\documentclass[10pt, journal, compsoc]{IEEEtran}
\IEEEoverridecommandlockouts

\usepackage[oldenum,olditem]{paralist}
\usepackage{comment}
\usepackage{amsmath,amssymb,amsfonts}
\usepackage{gensymb}
\usepackage{algorithmic}
\usepackage{graphicx}
\usepackage{textcomp}
\usepackage{multirow}
\usepackage[table]{xcolor}  
\usepackage{xcolor}
\usepackage{units}
\usepackage{color}
\usepackage{subfigure}
\usepackage{cite}
\usepackage{color,soul}
\usepackage[linesnumbered,ruled,vlined]{algorithm2e}
\usepackage{pifont}
\usepackage{mathtools}
\DeclarePairedDelimiter\ceil{\lceil}{\rceil}

\usepackage{multirow}
\usepackage[normalem]{ulem}
\usepackage[symbol]{footmisc}

\begin{document}

\title{PlaceRAN: optimal placement of virtualized network functions in the next-generation radio access networks}

\author{Fernando Zanferrari Morais, 
    Gabriel Matheus F. de Almeida, 
    Leizer Pinto, \\ 
    Kleber Vieira Cardoso, 
    Luis M. Contreras, 
    Rodrigo da Rosa Righi, and 
    Cristiano Bonato Both
\IEEEcompsocitemizethanks{
\IEEEcompsocthanksitem Fernando Zanferrari Morais, Rodrigo da Rosa Righi, and Cristiano Bonato Both are with the University of Vale do Rio dos Sinos (UNISINOS).
\IEEEcompsocthanksitem Gabriel Matheus, Leizer Pinto, Kleber Vieira Cardoso are with the Universidade Federal de Goiás (UFG).
\IEEEcompsocthanksitem Luis M. Contreras is with the Transport \& IP Networks - Systems and Network Global Direction, Telefónica CTIO Unit.
}
}

\markboth{IEEE Transactions on Mobile Computing,~Vol.~XX, No.~Y, MONTH~YEAR}%
{Shell \MakeLowercase{\textit{et al.}}: Bare Demo of IEEEtran.cls for IEEE Journals}

\maketitle

\IEEEpeerreviewmaketitle
\begin{abstract}
The fifth-generation mobile evolution enables several transformations on Next Generation Radio Access Networks (NG-RAN). The RAN protocol stack is splitting into eight possible disaggregated options combined into three network units, i.e., Central, Distributed, and Radio. Besides that, further advances allow the RAN software to be virtualized on top of general-purpose vendor-neutral hardware, dealing with the concept of virtualized RAN (vRAN). The disaggregated network units initiatives reach full interoperability based on the Open RAN (O-RAN). The combination of NG-RAN and vRAN results in vNG-RAN, enabling the management of disaggregated units and protocols as a set of radio functions. The placement of these functions is challenging since the best decision can be based on multiple constraints, such as the RAN protocol stack split, routing paths of transport networks with restricted bandwidth and latency requirements, different topologies and link capabilities, asymmetric computational resources, etc. This article proposes the first exact model for the placement optimization of radio functions for vNG-RAN planning, named PlaceRAN. The main objective is to minimize the computing resources and maximize the aggregation of radio functions. The PlaceRAN evaluation considered two realistic network topologies. Our results reveal that the PlaceRAN model achieves an optimized high-performance aggregation level, it is flexible for RAN deployment overcoming the network restrictions, and it is up to date with the most advanced vNG-RAN design and development.

\end{abstract}

\section{Introduction}
\label{intro}




\IEEEPARstart {T} {he} fifth-generation mobile evolution is based on standards \cite{3gpp.23.501,gstr2018transport,3gpp2017study}, enabling the functional radio protocol stack disaggregation and the virtualization of the Radio Access Network (vRAN). The instantiation of virtualized parts of such radio stack on general-purpose vendor-neutral hardware. These standards specify the Next Generation RAN (NG-RAN) architecture to meet the new service demands, e.g., ultra-low latency and high-bandwidth applications. Furthermore, the industry is leveraging on NG-RAN promoting open solutions, being Open RAN (O-RAN)\footnote{https://www.o-ran.org/} the most promising one, focused on interoperability among vendors’ implementations of the disaggregated radio protocol stack \cite{gavrilovska2020cloud}. The proposed NG-RAN architecture enables the base station for splitting the radio protocol stack into eight potential options combined in up to three network elements: (i) Central Unit (CU), (ii) Distributed Unit (DU), and (iii) Radio Unit (RU) \cite{gstr2018transport,marsch20185g}. These split options intend to enable sophisticated radio features and improve cost efficiency compared with previous mobile generations \cite{agrawal2017cloud,marsch20185g}.




The softwarization process in NG-RAN is guided by the virtualization of nodes and radio functions enabled by the Network Function Virtualization (NFV) concept \cite{badulescu2019etsi,bernardos2019network}. For example, the NG-RAN architecture functional split combined with vRAN provides flexibility in deploying mobile access networks. This flexibility allows mobile network operators to place the radio functions taking into account available network resources and existing user demand. The dynamic placement of radio functions following a fine-grained network management approach is vital for fifth-generation networks to achieve societal digital transformation expectations. However, the development of the virtualized NG-RAN architecture (vNG-RAN) is an unprecedented challenging problem since crosshaul transport networks (i.e., integrated backhaul, midhaul, and fronthaul networks) have stringent bandwidth and latency requirements, different diverse topologies and link capabilities, differently sized Computing Resources (CR), and unbalanced user demand \cite{gstr2018transport,sehier2019transport}. 

vNG-RAN represents the latest advancement in fifth-generation research. However, no decision-making methods are yet available for the design and specification of the placement of virtualized radio network functions. The placement is defined as an optimization problem \cite{masdari2016overview,laghrissi2018survey} based on the best joint decision between the split of the RAN protocol stack, the routing paths of the crosshaul network, and the CRs strategies of the CU, DU, and RU nodes \cite{laghrissi2018survey}. Therefore, the ideal placement leads to an analysis comprising the bandwidth and latency requirements for each split option between CU-DU and DU-RU. Moreover, each split option results in a computing cost (in terms of processing, memory, and storage) to be evaluated. The placement needs to be aware of the load occupation and the networking and computing resources allocation to ensure proper network scalability.

In the literature, several works address the placement optimization of radio functions. The main strategies developed so far are to maximize the number of Virtual Network Functions (VNFs) running in a single CU, DUs fixes, and close to RUs \cite{murti2020optimization, murti:20}. Moreover, CU is co-located with the core of the network \cite{garcia2018fluidran}. The state-of-the-art is restricted in considering the number of protocol disaggregation options, with a maximum of five \cite{fonseca:19}, or the overall efficiency obtained under crosshaul constraints (mainly, the fronthaul network) \cite{yusupov2018multi,murti2020optimization,harutyunyan2020cu}, and computing resources \cite{song2019clustered,arouk2017multi}. Therefore, to the best of our knowledge, there is no work yet in the literature fully considering CUs, DUs, and RUs on realistic operational networks, making the problem more general with higher functional split options and protocol stack analysis.

\textbf{Contributions}. In this article, we introduce PlaceRAN, a problem formulation for the optimal placement of vNG-RAN functions. The problem is formulated as the best trade-off between maximizing the aggregation level of virtualized NG-RAN functions and minimizing the number of computing resources necessary for running these functions. PlaceRAN innovates by considering in the formulation all the disaggregated RAN elements (CU, DU, and RU), the segments between those elements (fronthaul, midhaul, and backhaul), and also all possible functional splits according to the standards. Our contributions can be summarized as follows:


\begin{itemize}
    
    \item \textbf{New problem formulation} -- PlaceRAN is the most general problem formulation in the context of vNG-RAN, and it was designed with a comprehensive set of real-world NG-RANs considerations in mind.

    \item \textbf{New approach} -- we introduced some concepts to properly formulate PlaceRAN, such as Disaggregated RAN Combination (DRC) and multi-stage problem formulation, turning the problem formulation simple despite its generality.

    \item \textbf{Efficient exact solution} -- we solve PlaceRAN using a conventional solver (i.e., IBM CPLEX) for real-world RAN instances despite the problem complexity.
    
     \item \textbf{Evaluation and new insights} -- our evaluation used examples of present and future RANs. We show how PlaceRAN can contribute to the virtualization of present RANs, but it is also ready to deal with the optimal placement of forthcoming vNG-RANs.
     
    
    
    
\end{itemize}

\textbf{Article organization.} Section \ref{sec:concepts} introduces the background for vNG-RAN placement. The PlaceRAN system model and problem statement are described in Section \ref{sec:model}. Next, Section \ref{evaluation} presents the PlaceRAN evaluation methodology and results. The related work is discussed in Section \ref{rw}, and finally, Section \ref{conc} presents the final remarks.


\section{Virtualized NG-RAN Placement}
\label{sec:concepts}

The fundamental idea of a disaggregated NG-RAN is to decompose the RAN functions into virtualized components that can be distributed to run into different computing devices, i.e., following a non-monolithic approach in contrast to traditional solutions. Therefore, it is necessary to identify how this decomposition can be performed and which conditions must be satisfied to have the disaggregated version running correctly. The disaggregated NG-RAN is defined by the concept of functional splits that specifies all the possible partitions for the radio network functions, stating clear interface points between them, and the requirements for each of those radio network functions \cite{larsen:19,gavrilovska2020cloud}.

\begin{table}[!h]
   \caption{3GPP Latency and bitrate requirements for each split
    \cite{3gpp2017study}.}
    \label{tab:splits}
    \centering{ {
        \begin{tabular}{|c|l|c|c|c|}
        \hline
    Split & \multicolumn{1}{c|}{Functional} & One-way & \multicolumn{2}{c|}{Bitrate (Gbps)} \\ 
    \cline{4-5}
    Option & \multicolumn{1}{c|}{Split} & latency & DL & UL \\ \hline
    
    \rowcolor[HTML]{cdd5e4}  O1 &  RRC-PDCP      & 10 ms      & 4    & 3  \\
    
    O2 & PDCP - High RLC  & 10 ms      & 4    & 3  \\
    
    \rowcolor[HTML]{cdd5e4} O3 & High RLC -  Low RLC & 10 ms      & 4    & 3  \\ 
    
    O4 & Low RLC - High MAC  & 1 ms      & 4  & 3  \\ 
    
    \rowcolor[HTML]{cdd5e4} O5 & High MAC - Low MAC & $<$ 1 ms   & 4  & 3  \\ 
    
    O6 & Low MAC - High PHY & 250 $\mu$s & 4.13  & 5.64  \\ 
    
    \rowcolor[HTML]{cdd5e4} O7 & High PHY - Low PHY & 250 $\mu$s & 86.1$^*$  & 86.1$^*$  \\ 
    
    O8 & Low PHY - RF & 250 $\mu$s & 157.3 & 157.3 \\ \hline
    
       \end{tabular}
       O7 split maximum value.$^*$ 
    } }
\end{table}

    
    
    
    
    
    
    
    

\begin{figure*}[h!] 
 \begin{center}
\includegraphics[width=1.\textwidth]{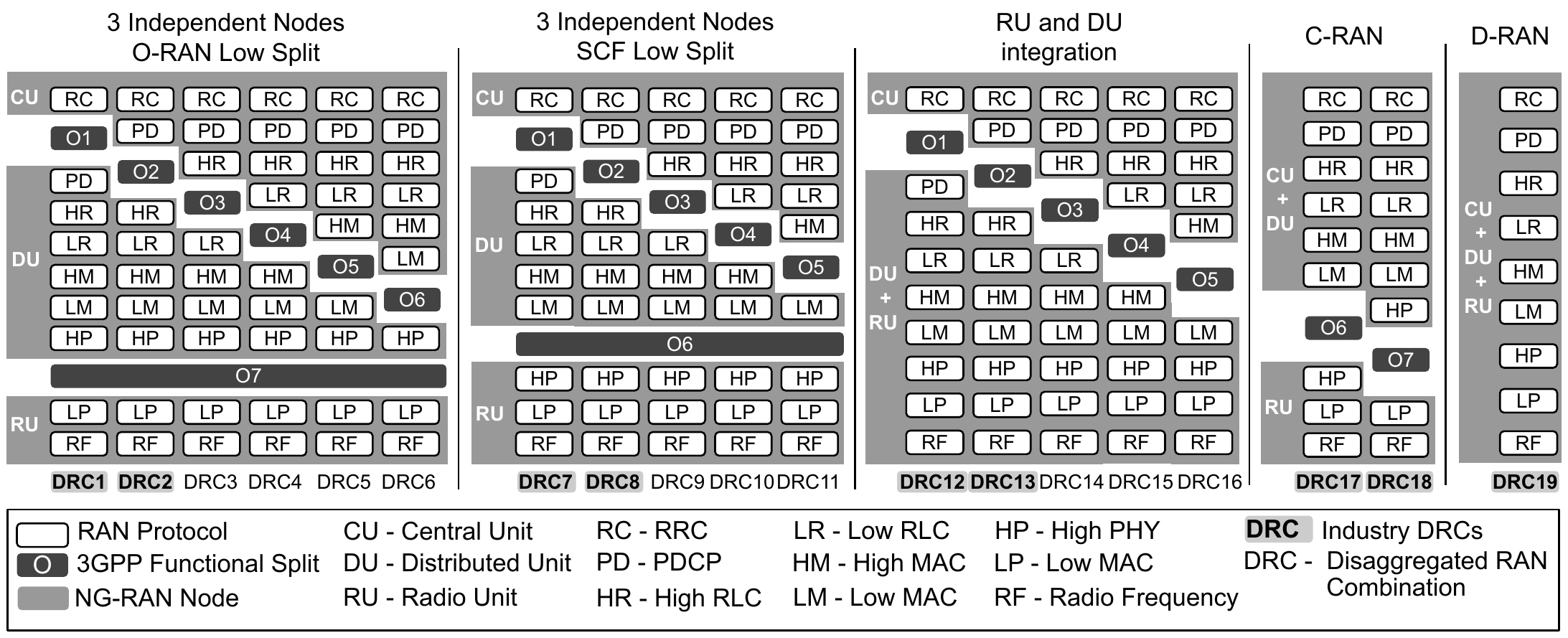}
  \end{center}
\caption{Functional split and computing devices for a disaggregated NG-RAN.}
\label{fig:DRC}
\end{figure*}



The number of functional splits and the partitions of the radio stack proposed are determined by specifications from standardization bodies, such as Release 14 from 3GPP \cite{3gpp2017study} and IMT-2020/5G from ITU-T \cite{gstr2018transport}. Table~\ref{tab:splits} shows the specifications of the disaggregated protocol stack considered in this work. The (maximum) latency and (minimum) bitrate values must be satisfied in the communication among the RAN nodes (CU, DU, and RU) even if they are running on different computing devices. The latency and bitrate must be assured according to the functional split specified. The (maximum) latency and (minimum) bitrate values presented in the table correspond to an RU with the following configuration: 100 MHz bandwidth of spectrum, 32 antenna ports, 8 MIMO layers, and 256 QAM modulation \cite{3gpp2017study}.

Each RAN node (CU, DU, and RU) is considered a virtualized network function as part of a disaggregated NG-RAN (3 independents nodes). Each of them runs different parts of a given functional split. Therefore, each RAN node may be identified by the set of the protocols running into it, as shown in Fig~\ref{fig:DRC}. A configuration with less than three nodes may be named DU and RU integration (DU and RU), C-RAN (CU and DU), or D-RAN (CU, DU, and RU) \cite{gstr2018transport}. In a disaggregated NG-RAN, the paths along the network connecting the core to vCU, vCU to vDU, and vDU to RU are defined as backhaul, midhaul, and fronthaul, respectively. This terminology is useful since each physical link of the access network acts as a crosshaul, meaning that it can transport any combination of the previous paths. The crosshaul needs to ensure the latency and bitrate required according to the variety of functional splits \cite{sehier2019transport,larsen:19}.

As illustrated in Fig.~\ref{fig:DRC}, each functional split and the corresponding placement of VNFs in a specific RAN node characterize a Disaggregated RAN Combination (DRC). The concept of DRC, introduced in this article, represents the preservation of the protocol stack order during the processing of VNFs. Nineteen DRCs are mapped considering seven split options. The O8 option cannot be virtualized since the RF protocol is hardware-based, making virtualization not feasible. Certain DRCs are not used in practice because they are restrictive in the limitations of crosshaul networks, e.g., midhaul with less than 1 ms.




We highlighted the nine DRCs effectively adopted in vRAN deployments, whose choice is based on standardization bodies and industry alliances \cite{3gpp.23.501, gstr2018transport, 3gpp2017study}. In both architectures of three independent nodes (O-RAN and SCF), the focus is on O1 and O2 (Fig.~\ref{fig:DRC} -  DRC1, DRC2, DRC7, and DRC8). The split O1 makes possible the decentralized data plane. 
Split O2 is consolidated by 3GPP and ITU-T via the F1 interface and is an industry reference for O-RAN and SCF initiatives \cite{gstr2018transport,mavenir,o-ran1}. Two industry DRCs were chosen for the DU and RU integration (Fig.~\ref{fig:DRC} - DRC12 and DRC13), besides the two C-RAN options (Fig.~\ref{fig:DRC} - DRC17 and DRC18). These splits align with ITU-T (mainly due to crosshaul constraints) and following O-RAN and SCF initiatives \cite{gavrilovska2020cloud,mavenir}. Naturally, the traditional D-RAN architecture is also supported to provide scenarios where the crosshaul is very limited \cite{gstr2018transport} (Fig.~\ref{fig:DRC} - DRC19).

In summary, the disaggregated NG-RAN can be implemented as a virtualized network service, i.e., a collection of VNFs with a particular set of characteristics. First, the service consists of the process of the full protocol stack per RF device in NG-RAN. This processing implies respecting an appropriate order of the flow-through VNFs, i.e., Service Function Chain (SFC). VNFs are instantiated in RAN nodes, which are also virtual elements that can run in different computing devices in NG-RAN. The choice of where to position RAN nodes and their VNFs affects the resources applied, including computing and networking. For each NG-RAN topology and set of resources, there may be multiple options for positioning VNFs and RAN nodes. In general, the objective is to consume the minimum resources and group the maximum of VNFs related to the same protocol or layer. However, each positioning option implies different computing and networking demands, which must not exceed the overall available resources. Therefore, the function’s placement becomes a complex optimization problem that we will formally present in the next section.

\section{Model and problem statement}\label{sec:model}


Initially, Subsection 3.1 presents the system model of a formally virtualized and disaggregated NG-RAN, in which different functional splits are possible. Moreover, the placement of the virtual functions is also introduced considering multiple options. After, Subsection 3.2 formulates the optimization problem to minimize the number of necessary computing resources, select the functional splits, and place the virtual functions.


\subsection{System Model}

\begin{figure}[htb] 
 \begin{center}
\includegraphics[width=.5\linewidth]{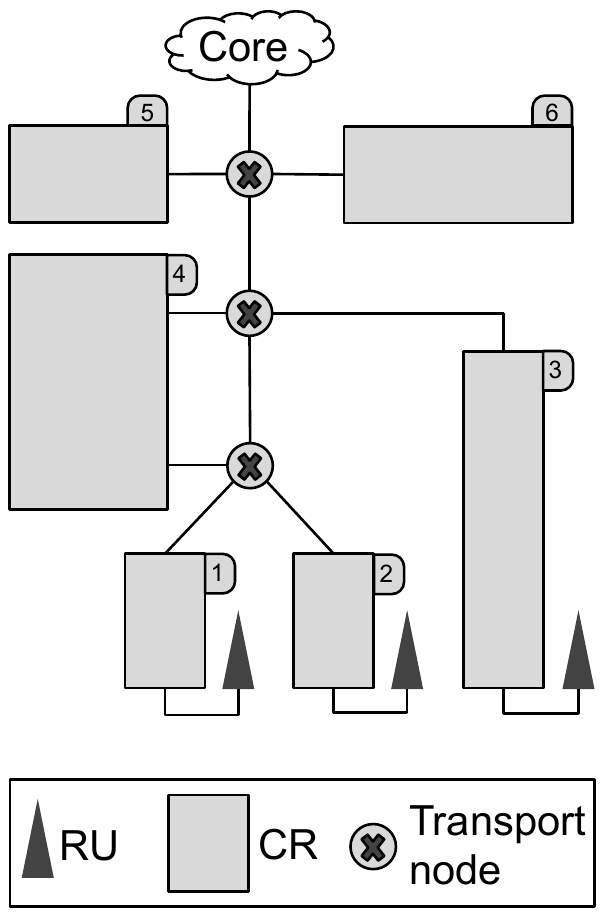}
  \end{center}
\caption{RAN considered as a reference to the system model.}
\label{fig:system_model}
\end{figure}

According to the 3GPP standards (Release 15 \cite{3gpp.23.501} and Release 16 \cite{3gpp.21.916}), we consider the RAN domain of a mobile network and its connection to the core network, as illustrated by Fig.~\ref{fig:system_model}. The RAN domain is composed of:

\begin{itemize}
    \item A set $\mathcal{B} = \{b_{1}, b_{2}, ..., b_{|\mathcal{B}|}\}$ of RUs, i.e., nodes hosting the Low PHY sublayer and the RF processing based on a lower layer functional split.
    \item A set $\mathcal{C} = \{c_{1}, c_{2}, ..., c_{|\mathcal{C}|}\}$ of CRs that may process the virtual functions. Each CR $c_m$ has a processing capacity $c^{Proc}_m$ (given as some reference number of cores). Moreover, each CR has other characteristics, such as memory and storage capacity, but they are not commonly exhausted before the processing capacity in the context of disaggregated RAN. A CR may connect directly to an RU.
    \item A set $\mathcal{T} = \{t_{1}, t_{2}, ..., t_{|\mathcal{T}|}\}$ of transport nodes, which may connect to RUs, CRs, core, or each other.
\end{itemize}

To represent RAN and core, we define the graph $G = (\mathcal{V}, \mathcal{E})$, with $\mathcal{V} = \{v_{0}\} \cup \mathcal{B} \cup \mathcal{C} \cup \mathcal{T}$ being the set of nodes and $\mathcal{E} = \{e_{ij};v_i,v_j\in V\}, \ v_{i},v_{j} \in \mathcal{V}\}$ representing the set of network links connecting the nodes. $v_{0}$ represents the core and it is the source/destination for all DRCs. Each link $e_{ij} \in \mathcal{E}$ has a transmitting capacity $e^{Cap}_{ij}$ (given in multiples of bps) and a latency $e^{Lat}_{ij}$ (given in fractions of a second).


\noindent \textbf{Paths and routing} -  we consider that all network traffic has the core as its source (downlink) or destination (uplink). However, without loss of generality, we represent only the downlink case in this work. We define $\mathcal{P}_{l}$ as the set of $k$-shortest paths from the core to each RU $b_{l} \in \mathcal{B}$. Each path $p \in \mathcal{P}_{l}$ is composed of three sub-paths: $p_{BH}$ (backhaul), $p_{MH}$ (midhaul), and $p_{FH}$ (fronthaul), in which at least one of these sub-paths is not empty.

\noindent \textbf{Virtualized RAN functions} - we consider that a VNF runs parts of the RAN protocol stack (except the RF protocol, as detailed previously). Moreover, VNFs are labeled in increasing order, starting from PHY Low with $f_1$ and ending at RRC with $f_8$. We define $\mathcal{F} = \{f_1, f_2, f_3, f_4, f_5, f_6, f_7, f_8\}$ as the set of disaggregated RAN VNFs, where the distribution must follow one of the industry DRCs \cite{o-ran1,gstr2018transport} of the set $\mathcal{D} = \{D_{1}, D_{2}, ..., D_{|\mathcal{D}|}\}$ (as illustrated in Fig.~\ref{fig:DRC}).

\subsection{PlaceRAN Problem Formulation}

Our problem formulation has two objectives: (i) maximize the aggregation level of RAN VNFs and (ii) minimize the number of CRs used for this aggregation. Since the computing and network capacities are limited, decreasing the number of CRs may not imply an increase in the aggregation level, which creates conflicting objectives. However, there is a relationship between the number of CRs and the aggregation level. Additionally, the functional splits' aggregation level is not measured only by the number of VNFs and CRs. The aggregation level is also affected by two other metrics: the number of DRCs employed and the priority or preference of each DRC. Since three incompatible metrics measure the aggregation level, we designed our formulation into three stages. The optimal solution can be eventually obtained in the first or second stage, but certainty is only obtained at the third stage, after resolving all potential draws.


\subsection*{First Stage}

In the first stage, the objective is to jointly maximize the number of grouped RAN VNFs and minimize the number of CRs used for running those VNFs. We define $x^{p,r}_{l} \in \{0, 1\}$ as the decision variable representing which pair of path $p \in \mathcal{P}_{l}$ and DRC $D_r \in \mathcal{D}$ is selected to serve RU $b_l \in \mathcal{B}$. From the input data, we determine $u^{p}_{m} \in \{0, 1\}$ to indicate if $c_m \in \mathcal{C}$ is part of the $p \in \mathcal{P}_{l}$. Additionally, we define the mapping function $M(c_m, f_s, b_l, D_r) \in \{0, 1\}$ over the input data, which indicates if the CR $c_m \in \mathcal{C}$ runs the VNF $f_s \in \mathcal{F}$ from the RU $b_l \in \mathcal{B}$, according to the DRC $D_r \in \mathcal{D}$. Therefore, we define the following objective function:

\begin{equation}
\label{eq:of1}
    \textit{minimize} \ \ \ \ \ \Phi_1 - \Phi_2,    
\end{equation}

where $\Phi_1$ represents the amount of CRs, given by:

\begin{equation}
\Phi_1 =  \sum_{c_m \in C} \ceil*{\frac{\sum_{b_l \in B} \sum_{D_r \in \mathcal{D}} \sum_{p \in \mathcal{P}_l} \left ( x^{p,r}_{l} \cdot u^{p}_{m} \right )}{|\mathcal{C}|}},
\end{equation}

and $\Phi_2$ represents the amount of grouped RAN VNFs, given by:

\begin{equation}
\begin{split}
\Phi_2 =  & \sum_{c_m \in \mathcal{C}} \sum_{f_s \in \mathcal{F}} \\
& \left( \sum_{b_l \in \mathcal{B}} \sum_{D_r \in \mathcal{D}} \sum_{p \in \mathcal{P}_l} \left [ x^{p,r}_{l} \cdot u^{p}_{m} \cdot M(c_m, f_s, b_l, D_r) \right ] \right.\\
 & \left.- \ceil*{\frac{\sum\limits_{D_r \in \mathcal{D}} \sum\limits_{p \in \mathcal{P}_{l}} \sum\limits_{b_l \in \mathcal{B}} \left [ x^{p,r}_{l} \cdot u^{p}_{m} \cdot M(c_m, f_s, b_l, D_r) \right ]}{|\mathcal{F}|}} \right).
\end{split}
\end{equation}

For each RU $b_l \in \mathcal{B}$, exactly one DRC $D_r \in \mathcal{D}$, using a single path $p \in \mathcal{P}_{l}$, must be selected, as represented by the following constraint:

\begin{equation}
\label{eq:constraint1}
    \sum_{D_r \in \mathcal{D}} \sum_{p \in \mathcal{P}_{l}} x^{p,r}_{l} = 1, \ \ \ \ \ \ \ \ \forall b_l \in \mathcal{B}.
\end{equation}

The transmitting capacity $e^{Cap}_{ij}$ of every link $e_{ij}$ must not be exceeded, as described by the following constraint:

\begin{equation}
\label{eq:constraint2}
\begin{split}
    \sum_{b_l \in \mathcal{B}} \sum_{D_r \in \mathcal{D}} \sum_{p \in \mathcal{P}_{l}} \Big[  x^{p,r}_{l} \Big( & y^{p_{BH}}_{e_{ij}} \cdot \alpha^r_{BH} + y^{p_{MH}}_{e_{ij}} \cdot \alpha^r_{MH} + \Big. \Big.\\ 
    & \Big. \Big. y^{p_{FH}}_{e_{ij}} \cdot \alpha^r_{FH} \Big ) \Big ] \leq e^{Cap}_{ij},  \forall e_{ij} \in \mathcal{E},
\end{split}
\end{equation}

where $y^{p_{BH}}_{e_{ij}}$, $y^{p_{MH}}_{e_{ij}}$, and $y^{p_{FH}}_{e_{ij}}$ indicate if the link $e_{ij}$ is part of the backhaul, midhaul, or fronthaul, respectively, in a path $p \in \mathcal{P}_{l}$ that transports a specific DRC $D_r \in \mathcal{D}$. Each $D_r \in \mathcal{D}$ has associated demands for bitrate in the backhaul, midhaul, and fronthaul, represented by $\alpha^r_{BH}$, $\alpha^r_{MH}$, and $\alpha^r_{FH}$, respectively. There are functional splits in which the path $p \in \mathcal{P}_{l}$ has less than three sub-paths, e.g., DRCs 12, 17, and 19 (as illustrated in Fig.~\ref{fig:DRC}). Moreover, if the sub-path is absent, then any link is part of it.

Each $D_r \in \mathcal{D}$ tolerates a maximum latency in each sub-path (backhaul, midhaul, and fronthaul) of the path $p \in \mathcal{P}_{l}$, which is described by the following constraints:

\begin{equation}
\label{eq:constraint3}
    \sum_{e_{ij} \in \mathcal{E}} x^{p,r}_{l} \cdot y^{p_{BH}}_{e_{ij}} \cdot e^{Lat}_{ij} \leq \beta^r_{BH}, \forall b_l \in \mathcal{B}, p \in \mathcal{P}_{l}, D_r \in \mathcal{D},
\end{equation}

\begin{equation}
\label{eq:constraint4}
    \sum_{e_{ij} \in \mathcal{E}} x^{p,r}_{l} \cdot y^{p_{MH}}_{e_{ij}} \cdot e^{Lat}_{ij} \leq \beta^r_{MH}, \forall b_l \in \mathcal{B}, p \in \mathcal{P}_{l}, D_r \in \mathcal{D},
\end{equation}

\begin{equation}
\label{eq:constraint5}
    \sum_{e_{ij} \in \mathcal{E}} x^{p,r}_{l} \cdot y^{p_{FH}}_{e_{ij}} \cdot e^{Lat}_{ij} \leq \beta^r_{FH}, \forall b_l \in \mathcal{B}, p \in \mathcal{P}_{l}, D_r \in \mathcal{D},
\end{equation}

where $\beta^r_{BH}$, $\beta^r_{MH}$, and $\beta^r_{FH}$ represent the maximum latency tolerated in the backhaul, midhaul, and fronthaul, respectively, of a path $p \in \mathcal{P}_{l}$ that transports a specific DRC $D_r \in \mathcal{D}$. There are functional splits in which the path $p \in \mathcal{P}_{l}$ has less than three sub-paths. In the same way, if the sub-path is absent, then no link is part of it.

Finally, the VNFs selected to run in a CR $c_m \in \mathcal{C}$ must not exceed its processing capacity $c^{Proc}_{m}$, as represented by the following constraint:

\begin{equation}
\label{eq:constraint6}
\begin{split}
    \sum_{f_s \in \mathcal{F}} \sum_{b_l \in \mathcal{B}}
    \sum_{D_r \in \mathcal{D}} \sum_{p \in \mathcal{P}_{l}} &
    x^{p,r}_{l} \cdot u^{p}_{m} \cdot M(c_m, f_s, b_l, D_r)  \cdot \gamma^{s}_{m} \\
    & \leq c^{Proc}_{m}, \ \ \ \ \ \ \forall c_m \in \mathcal{C},
\end{split}
\end{equation}

where $\gamma^{s}_{m}$ is the computing demand of the VNF $f_s \in \mathcal{F}$.

\subsection*{Second Stage}

\begin{figure}[htb] 
 \begin{center}
\includegraphics[width=1.\linewidth]{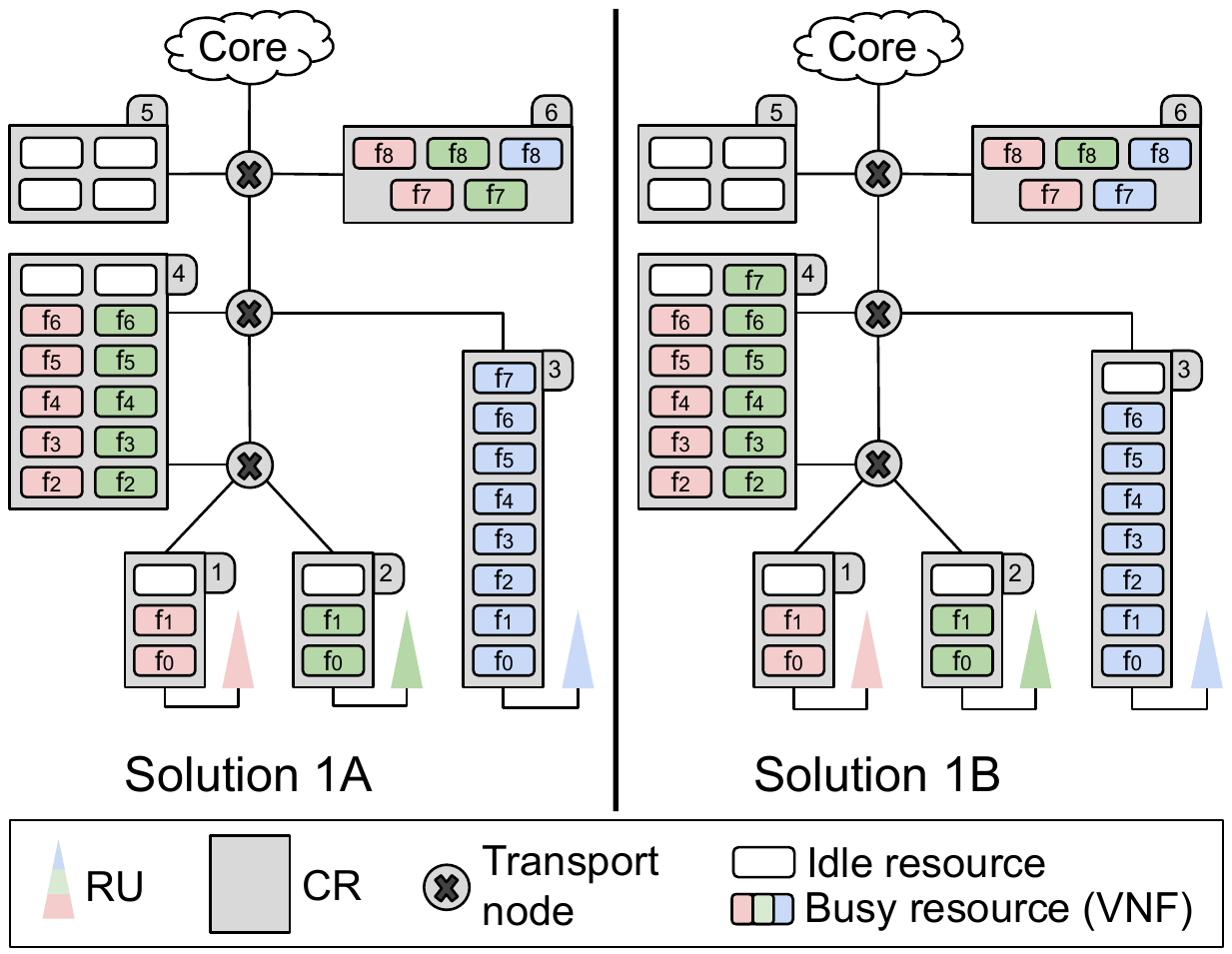}
  \end{center}
\caption{A possible tie after solving the first stage.}
\label{fig:stage1-a_tie}
\end{figure}

After solving the first stage, we obtain the minimum number of CRs necessary to achieve the maximum aggregation level of RAN VNFs. Since these two objectives may be conflicting, the final result of the first stage represents the best trade-off between these goals. However, the aggregation level achieved may not be optimal. An example is illustrated in Fig.~\ref{fig:stage1-a_tie}, where solution 1A and solution 1B are two possible solvers' outcomes. In this case, both solutions have the same value of object function, i.e., both achieve the same value for Equation~(\ref{eq:of1}).

While comparing different solutions with the same value for the objective function (Equation~(\ref{eq:of1})) in the first stage, we observed that those with a smaller number of unique DRCs combinations, sharing the same virtualized RAN function $\mathcal{F}$ and CR, improve the system performance. For example, in Fig.~\ref{fig:stage1-a_tie}, there is a benefit in sharing CR 6 by $f_7$ (PDCP) from the red DRC and $f_7$ from the green DRC (solution 1A). However, there is no benefit in sharing CR 6 by $f_7$ (PDCP) from the red DRC and $f_7$ from the blue DRC (solution 1B). Solution 1A has two unique DRCs: 2 and 12, while solution 1B has three unique DRCs: 1, 2, and 13. This example illustrates the need to optimize the overall amount of DRCs. Therefore, the objective function of the second stage is to minimize the number of unique DRCs:

\begin{equation}
\label{eq:of2}
    \textit{minimize} \ \ \ \ \ \sum_{D_r \in \mathcal{D}} \ceil*{\frac{\sum_{b_l \in \mathcal{B}} \sum_{p \in \mathcal{P}_{l}} x^{p,r}_{l}} {|\mathcal{B}|}}.
\end{equation}

This second stage must consider only solutions with exactly the same value of the objective function achieved by the optimal solution at the first stage. The following constraint assures this situation:

\begin{equation}
\label{eq:of2-c1}
    \Phi_1 - \Phi_2 = f_{1st\_stage}(x^{p,r*}_{l}),    
\end{equation}

where $f_{1st\_stage}(x^{p,r*}_{l})$ represents the objective function's value from the first stage when the optimal solution is found. Additionally, all constraints from the first stage must also be satisfied, i.e., the second stage is subject to the constraints (\ref{eq:constraint1})--(\ref{eq:constraint6}).

\subsection*{Third Stage}

\begin{figure}[htb] 
 \begin{center}
\includegraphics[width=1.\linewidth]{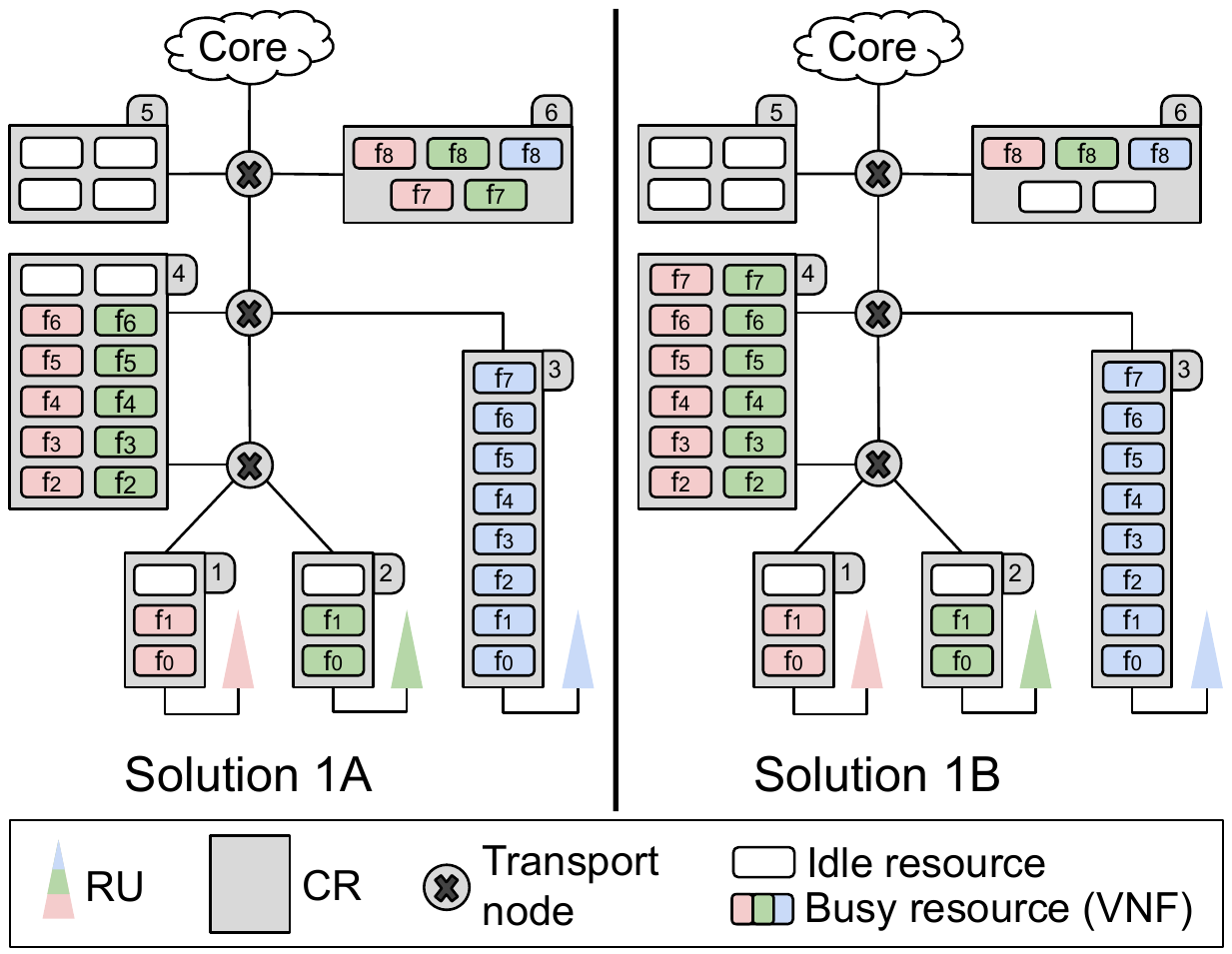}
  \end{center}
\caption{A possible tie after solving the second stage.}
\label{fig:stage2-a_tie}
\end{figure}

After solving the second stage, we eliminate potential solutions that have incompatible layers sharing common CRs. While this improves the aggregation level, there is still the possibility of obtaining different solutions with the same number of DRCs, but not functionally equivalent. An example is illustrated in Fig.~\ref{fig:stage2-a_tie}, where solution 2A and solution 2B are two possible solvers' outcomes. These solutions have the same value of object function, i.e., achieving the same value for Equation~(\ref{eq:of2}). In this case, solution 2A has two (unique) DRCs (2 and 12), while solution 2B also has two (unique) DRCs (1 and 12).

The solutions in Fig.~\ref{fig:stage2-a_tie} are not equivalent because each DRC has characteristics that make it unique, i.e., it is possible to rank DRCs and the solution's quality. By ranking DRCs, we are also able to differentiate again the solutions. Such ranking can be directly taken from the standard specifications, which already specify the preference order of DRCs. Naturally, different standardization bodies (e.g., O-RAN and ITU) may assign other priorities to DRCs, but our model is generic and works properly with any of them. In the priority assignment for each DRC, the smaller the value, the higher priority. Therefore, the objective function of the third stage is to minimize the sum of values assigned to DRCs:

\begin{equation}
    \textit{minimize} \ \ \ \ \ \sum_{b_l \in \mathcal{B}} \sum_{D_r \in \mathcal{D}} \sum_{p \in \mathcal{P}_{l}} \left (x^{p,r}_{l} \cdot D_{r}^{\omega} \right ),
\end{equation}

where $D_{r}^{\omega}$ represents the priority of the DRC $D_{r}$.

The third stage must also consider only solutions with exactly the same value of the objective function achieved by the optimal solution ay both the first stage, i.e., the constraint described by Equation~\ref{eq:of2-c1}, and the exact value of the objective function given by the optimal solution of the second stage. The following constraint assures this definition:


\begin{equation}
    \sum_{D_r \in \mathcal{D}} \ceil*{\frac{\sum_{b_l \in \mathcal{B}} \sum_{p \in \mathcal{P}_{l}} x^{p,r}_{l}} {|\mathcal{B}|}} = f_{2nd\_stage}(x^{p,r*}_{l}),  
\end{equation}

where $f_{2nd\_stage}(x^{p,r*}_{l})$ represents the value of the objective function from the second stage when the optimal solution is found. Similarly to the second stage, all constraints from the first stage must also be satisfied, i.e., the third stage is subject to the constraints (\ref{eq:constraint1})--(\ref{eq:constraint6}).



\section{Evaluation}
\label{evaluation}

This section evaluates the PlaceRAN model in several RAN configurations, including T1 and T2 as present and future network topologies, different amounts of resources, and distinct demands. Subsection 4.1 provides a general description of the evaluated scenarios, detailing the topologies and resources, varying parameters to assess the solution results. Subsection 4.2 presents and discusses the results obtained in the evaluation of PlaceRAN, which involves minimization of CRs versus maximization aggregations level, DRC selection, general aspects of the solutions, and characteristics of the optimization model.

\subsection{Description of the Scenarios}

Table \ref{tab:evalparam2} summarizes the scenarios employed in the evaluation of the PlaceRAN model. There are three types of scenarios: Low Capacity (LC), Random Capacity (RC), and High Capacity (HC). Each scenario has four types of Transport Nodes, which are defined according to the proximity to the Core and the number of neighbors: aggregation node 1 (AG1), aggregation node 2 (AG2), access node 1 (AC1), and access node 2 (AC2). Based on each transport node, there are four characteristics of real networks: (i) number of CRs, (ii) Bandwidth, (iii) Latency, and (iv) number of RU nodes. Moreover, our evaluation is focused on comparing the scenarios using two topologies. One on hand, T1 represents a current RAN deployment based on the 5G-crosshaul project \cite{5G-crosshaul} (in Table \ref{tab:evalparam2} as ${\dagger}$). On the other hand, T2 shows a trend in the design of the future RAN topologies, being aligned with the analysis performed in the PASSION project \cite{passion} (in Table \ref{tab:evalparam2} as ${\ddagger}$ ).


\begin{table*}[!h]
    \caption{Scenarios employed in the evaluation.}
    \label{tab:evalparam2}%
    \centering{ {
        \begin{tabular}{|ll|c|c|c|c|c|c|c|c|c|c|c|c|}
        \hline

        \multicolumn{2}{|c|}{Scenarios} & \multicolumn{4}{c}{Low Capacity (LC)} & \multicolumn{4}{|c}{Random Capacity (RC)} & \multicolumn{4}{|c|}{High Capacity (HC)}  \\ \hline
        
        \rowcolor[HTML]{cdd5e4} 
        \multicolumn{2}{|l|}{Transport Nodes} & AG1 & AG2 & AC1 & AC2 & AG1 & AG2 & AC1 & AC2 & AG1 & AG2 & AC1 & AC2 \\  \hline
        
      \multicolumn{1}{|l}{\multirow{1}{*}{Computing}} &
        & & & & &
      \shortstack{16\\ 32} &
      \shortstack{16\\ 32} &
      \shortstack{8\\ 16} &
      \shortstack{8\\ 16} &
      \multirow{-2}{*}{32} &
      \multirow{-2}{*}{32} &
      \multirow{-2}{*}{16} &
      \multirow{-2}{*}{16} \\ \cline{7-14}
      
      \multicolumn{1}{|l}{\multirow{-3}{*}{Resources}} &
      \multirow{-4}{*}{\shortstack{\hspace{0.5cm} $\dagger$ \\ \\ \\\hspace{0.5cm} $\ddagger$}} &
      \multirow{-4}{*}{16} &
      \multirow{-4}{*}{16} &
      \multirow{-4}{*}{8} &
      \multirow{-4}{*}{8} & 
      \shortstack{16\\ 64} &
      \shortstack{16\\ 64} &
      \shortstack{8\\ 32} &
      \shortstack{8\\ 32} &
      \multirow{-2}{*}{64} &
      \multirow{-2}{*}{64} &
      \multirow{-2}{*}{32} &
      \multirow{-2}{*}{32} \\ \hline 
      
      \rowcolor[HTML]{cdd5e4}
      & & 
      \multirow{-2}{*}{100} &
      \multirow{-2}{*}{40} &
      \multirow{-2}{*}{25} &
      \multirow{-2}{*}{10} &
      \shortstack{100\\ 400} &
      \shortstack{40\\ 100} &
      \shortstack{25\\ 40} &
      \shortstack{10\\ 25} &      
      \multirow{-2}{*}{400} &
      \multirow{-2}{*}{100} &
      \multirow{-2}{*}{40} &
      \multirow{-2}{*}{25} \\ \cline{3-14} \cline{3-14} \cline{3-14}

      \rowcolor[HTML]{cdd5e4} 
      \multirow{-4}{*}{Bandwidth (Gbps)} &
      \multirow{-4}{*}{\shortstack{\hspace{0.5cm} $\dagger$ \\ \\ \\ \hspace{0.5cm} $\ddagger$}} &
      \multirow{-2}{*}{800} &
      \multirow{-2}{*}{100} &
      \multirow{-2}{*}{50} &
      \multirow{-2}{*}{40} &
      \shortstack{1000\\ 800} &
      \shortstack{400\\ 100} &
      \shortstack{100\\ 50} &
      \shortstack{50\\ 40} &      
      \multirow{-2}{*}{1000} &
      \multirow{-2}{*}{400} &
      \multirow{-2}{*}{100} &
      \multirow{-2}{*}{50} \\ \hline 


      \multicolumn{1}{|c}{} & 
      \multicolumn{1}{l}{\hspace{-1.3cm} Computing \hspace{0.3cm}$\dagger,\ddagger$}  & 
      \multicolumn{4}{|c|}{0.002} &
      \multicolumn{4}{c}{0.05-0.002} &
      \multicolumn{4}{|c|}{0.05} \\ \cline{3-14}
        
      \multicolumn{1}{|c}{\hspace{-1.4cm} Latency} & 
      \multicolumn{1}{l}{\hspace{-1.3cm} Fiber \hspace{1.1cm}$\dagger,\ddagger$} & 
      \multicolumn{4}{|c|}{0.000005} & 
      \multicolumn{4}{c}{0.000005} & 
      \multicolumn{4}{|c|}{0.000005} \\ \cline{3-14}
        
      \multicolumn{1}{|c}{\hspace{-1.5cm} (ms)} & 
      \multicolumn{1}{l}{\hspace{-1.3cm} Optic \hspace{1cm}$\ddagger$} & 
      \multicolumn{4}{|c|}{0.010625} & 
      \multicolumn{4}{c}{0.010625} & 
      \multicolumn{4}{|c|}{0.010625} \\ \cline{3-14}
        
      \multicolumn{1}{|c}{} & 
      \multicolumn{1}{l}{\hspace{-1.3cm} Regenerator \hspace{0.17cm}$\ddagger$} & 
      \multicolumn{4}{|c|}{0.0005} & 
      \multicolumn{4}{c}{0.0005} & 
      \multicolumn{4}{|c|}{0.0005} \\ \hline

      \rowcolor[HTML]{cdd5e4}
      \multicolumn{2}{|l|}{RU Nodes \hspace{1.95cm}$\dagger,\ddagger$} & \multicolumn{1}{c|}{-} & \multicolumn{3}{c|}{F1 and R1} & 
      \multicolumn{1}{c|}{-} & \multicolumn{3}{c|}{F1 and R1} & 
      \multicolumn{1}{c|}{-} & \multicolumn{3}{c|}{F1 and R1} \\ \hline
      
      
      \end{tabular}
    } }
\end{table*}

\begin{figure}[h!] 
 \begin{center}
\includegraphics[width=1.\linewidth]{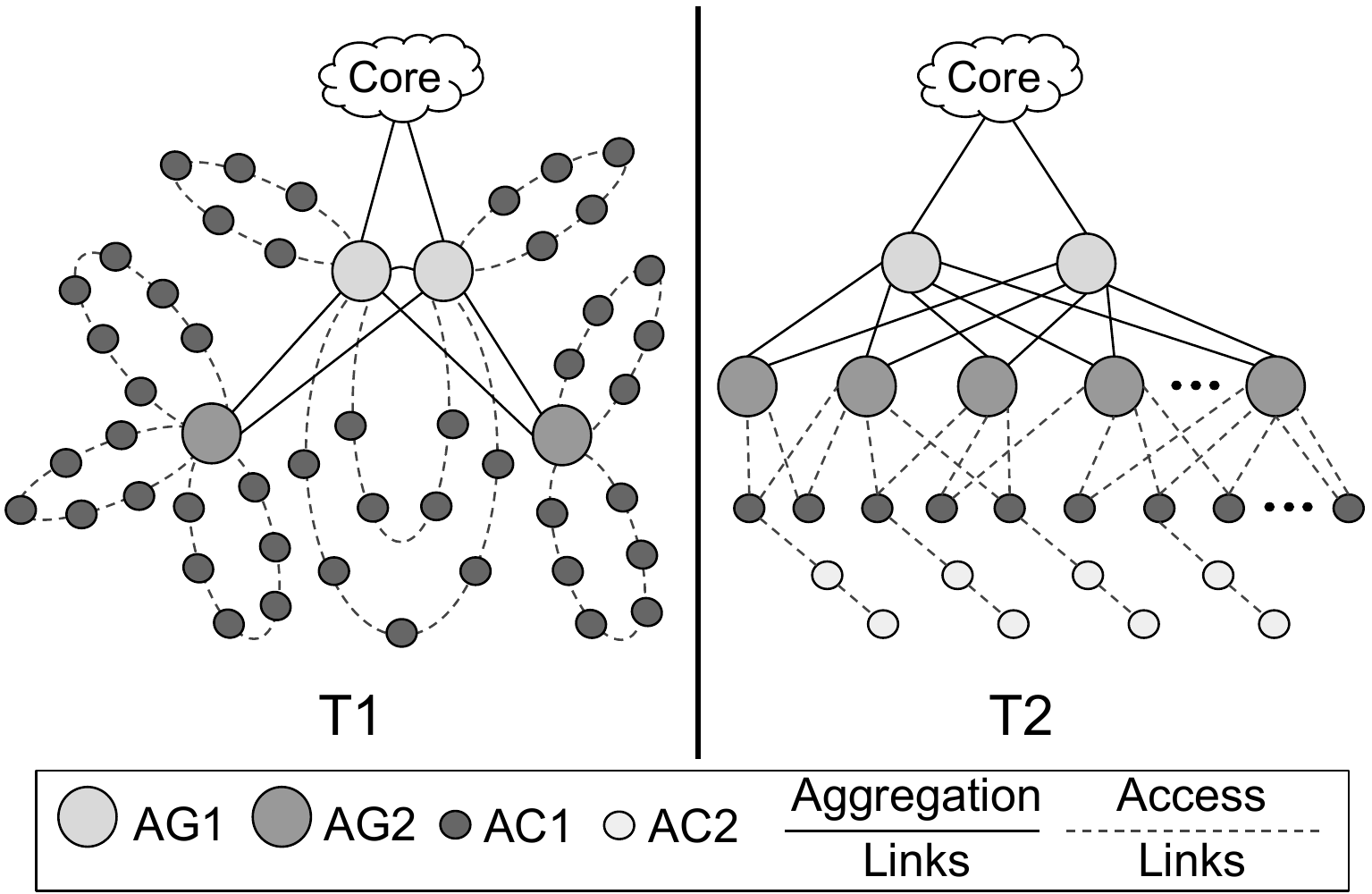}
  \end{center}
\caption{Types of evaluated RAN topology.}
\label{fig:topologies}
\end{figure}

\textbf{Transport Nodes} - we associate the nodes according to T1 and T2 topologies. T1 provides a current operational network of 51 nodes connected through a ring structure, formed by an aggregation ring and additional access rings. T2 proposes an expected trend in future RAN design with a hierarchical tree structure, presenting two aggregation stages and additional access stages. As suggested by the PASSION project, we limit T2 to 128 nodes. Fig. \ref{fig:topologies} illustrates the two RAN topologies employed in this evaluation. In both topologies, the transport nodes are classified into four types (AG1, AG2, AC1, and AC2\footnote[1]{There is no AC2 in the T1 topology due to the rings.}), as described and illustrated in the figure. On the one hand, this classification helps on having flexibility in the choice of parameters (i.e., the values of the characteristics) for each topology and scenario. Otherwise, the number of parameters is small (i.e., only four) in comparison to the number of transport nodes, e.g., 51 in the smallest topology. 



\textbf{Computing Resources} - we focused on the processing capacity (i.e., CPU) because this metric has been the most common bottleneck for computing devices in the context of network planning and vRAN optimization \cite{lin2018performance,hu2019architectural}. CR values in Table \ref{tab:evalparam2} indicate the number of CPUs (or a range of them in the case of RC scenario) that are directly connected to each type of transport node. In this RC scenario, a specific value is randomly selected (within the range) at the time of generating input data. The capacity of CRs is designed considering the CPU utilization profile of the RAN software as obtained from OpenAirInterface (OAI) implementation \cite{fujitsu}, as shown in Table \ref{tab:RANprotocol}. The exact values may vary according to the adopted software components and the computing device, but similar profiles have been reported in different works.

\begin{table}[!h]
    \caption{RAN Protocol Stack CPU Utilization \cite{yeoh2016performance,hu2019architectural,fujitsu}.}
  \label{tab:RANprotocol}%
    \centering{ {
        \begin{tabular}{|l|c|}
        \hline
        RAN Protocol & CPU Utilization (cores) \\ \hline
        
        \rowcolor[HTML]{cdd5e4} RRC & 0.49 \\
        
        PDCP & 0.49  \\
        
        \rowcolor[HTML]{cdd5e4} High RLC & 0.0245 \\
        
        Low RLC & 0.0245 \\
        
        \rowcolor[HTML]{cdd5e4}High MAC & 0.343 \\
        
        Low MAC & 0.343 \\
        
        \rowcolor[HTML]{cdd5e4} High PHY & 0.833 \\
        
        Low PHY & 2.352 \\    \hline
    
         \rowcolor[HTML]{3c4563} \color{white} Total & \color{white} 4.9 \\ \hline
       \end{tabular}
    } }
\end{table}

\textbf{Bandwidth} - the link capacity dimensioning considers the 5G-Crosshaul and PASSION projects and their strategies, i.e., current and future networks, respectively. We define the bandwidth following the standard bit rates provided by IEEE Alliance \cite{inteth}. In this sense, we distinguish the interface capacities between the two network topologies. 5G-Crosshaul uses links capacity from 40 Gbps to 400 Gbps in the aggregation nodes (AG1 and AG2) and 10 Gbps to 40 Gbps in the access nodes (AC1 and AC2). The PASSION Project (future network) operates from 100 Gbps to 1 Tbps in the aggregation nodes and from 40 Gbps to 100 Gbps in the access nodes.

\textbf{Latency} - to obtain the latency estimates, we considered four components: (i) the computing latency, referred as the time consumed in the forwarding process; (ii) the fiber latency, related to propagation delay in optical fibers; (iii) the transit delay of optic transmission, which is the delay of the optical device without any electronic processing; and, (iv) the delay dies to the signal regenerator, which transforms the optical signal into electric signals. Based on these four components, we employed two strategies, one for each topology. In the first strategy, T1 uses only the Computing and Fiber components. Since in T1 all link distances are available, the propagation delay is directly computed by the distance versus delay propagation. In the second strategy, based on T2, it is not possible to directly compute detailed propagation delay per link, although the rest of the components are considered. However, in this case, the PASSION project performed a study from where to derived statistical information about potential topologies minimum, average, and maximum link distance, and the number of hops. In consequence, for this analysis, it is assumed that the link distance applicable for the HC scenario, while for LC scenario average and maximum are taken, finally with RC scenario considering all of them \cite{huaweidwdm}.

\textbf{RU Nodes} - to evaluate the scenarios (and topologies) with different demands, we considered two configurations for the number of RU nodes connected to the transport nodes: F1 (Fixed 1) -- exactly one RU node is connected to each transport node and R1 (Random 1) -- zero or one (randomly chosen during input data generation) RU nodes are connected to each transport node. No RU node is connected to any AG1 in any topology scenario, as recommended by 5G-Crosshaul \cite{5G-crosshaul} and PASSION \cite{passion} projects. In our evaluation, the following number of RU nodes in each topology and configuration is assumed: 39 in T1-R1, 49 in T1-F1, 101 in T2-R1, and 126 in T2-F1. It is worth reminding that each RU node is assigned just one DRC.

\begin{table*}[!h]
    \caption{Parameters of the functional splits.}
    \label{tab:profiles}%
    \centering
    \begin{tabular}{|c|c|c|c|c|c|c|c|c|c|c|c|c|}
    \hline
    \multicolumn{2}{|c|}{DRC} &
    \multicolumn{2}{c|}{Split Options} &
    \multicolumn{3}{c|}{Tolerated latency - one way (ms)} &
    \multicolumn{3}{c|}{5G-Crosshaul bandwidth (Gbps)} &
    \multicolumn{3}{c|}{PASSION bandwidth (Gbps)} \\ \hline
    N$^o$ &
    Priority &
   High &
   Low &
   Core-CU &
   CU-DU &
   DU-RU &
   Core-CU &
   CU-DU &
   DU-RU &
   Core-CU &
   CU-DU &
   DU-RU \\ \hline
   
  \rowcolor[HTML]{cdd5e4} 1  & 4  & O1 & O7 & 1.5$\sim$10 & 1.5$\sim$10 & 0.250 & 2.97 & 5.4 & 17.4 & 9.9 & 13.2 & 42.6 \\ \hline

  2  & 1  & O2 & O7 & 1.5$\sim$10 & 1.5$\sim$10 & 0.250 & 2.97 & 5.4 & 17.4 & 9.9 & 13.2 & 42.6 \\ \hline

  \rowcolor[HTML]{cdd5e4} 7  & 6  & O1 & O6 & 1.5$\sim$10 & 1.5$\sim$10 & 0.250 & 2.97 & 5.4 & 5.6  & 9.9 & 13.2 & 13.6 \\ \hline

  8  & 5  & O2 & O6 & 1.5$\sim$10 & 1.5$\sim$10 & 0.250 & 2.97 & 5.4 & 5.6  & 9.9 & 13.2 & 13.6 \\ \hline

  \rowcolor[HTML]{cdd5e4} 12 & 10  & O1 & -  & 1.5$\sim$10 & 1.5$\sim$10 & -   & 2.97 & 5.4 & -    & 9.9 & 13.2 & -    \\ \hline

  13 & 9  & O2 & -  & 1.5$\sim$10 & 1.5$\sim$10 & -   & 2.97 & 5.4 & -    & 9.9 & 13.2 & -    \\ \hline

  \rowcolor[HTML]{cdd5e4}17 & 8  & -  & O6 & 1.5$\sim$10 & -           & 0.250 & 2.97 & -   & 5.6  & 9.9 & -    & 13.6 \\ \hline

  18 & 7 & -  & O7 & 1.5$\sim$10 & -           & 0.250 & 2.97 & -   & 17.4 & 9.9 & -    & 42.6 \\ \hline

  \rowcolor[HTML]{cdd5e4} 19 & 25 & -  & -  & 1.5$\sim$10 & -           & -   & 2.97 & -   & -    & 9.9 & -    & -    \\ \hline

  \end{tabular}%
\end{table*}

Table \ref{tab:profiles} summarizes the parameters of the functional splits employed in the evaluation. The priority of each DRC (used in the third stage of the PlaceRAN model) follows the O-RAN Alliance specifications \cite{o-ran1}. According to the split options, the maximum tolerated latency is defined for each RAN sub-path: backhaul (Core-CU), midhaul (CU-DU), and fronthaul (DU-RU). Moreover, according to the split options, the minimum acceptable bandwidth is defined for each RAN sub-path. However, we adopted different RAN configurations for T1 and T2 topologies, which implied the consideration of different bandwidths for each network, as shown in Table~\ref{tab:profiles}. In T1, we assumed the RAN configuration has the following characteristics: 40 MHz bandwidth, 32 antenna ports, 8 MIMO layers, 216 Physical Resource Blocks (PRBs), and 15 kHz subcarrier spacing per macro BS \cite{thor}. In T2, we assumed the RAN configuration has the following characteristics:  100 MHz bandwidth, 32 antenna ports, 8 MIMO layers, 132 PRBs, and 60 kHz subcarrier spacing per macro BS \cite{thor}.


We run all experiments in a Virtual Machine (VM) with Ubuntu 18.04, 16 vCPUs, 256 GB RAM, and 40 GB of the virtual disk. The VM is hosted in a server DELL PowerEdge M620 with two Intel Xeon E5-2650 @ 2 GHz. We used Python 2.7.17 and docplex 2.4.61 for implementing the PlaceRAN model, and the solver used was IBM CPLEX 12.8.0. The source code and the input data used in the evaluation are publicly available on Github\footnote{https://github.com/LABORA-INF-UFG/paper-FGLKLRC-2021}.

\subsection{Results}
\label{results}

We organize the evaluation of the PlaceRAN model into four parts as described in the following. The first part of the evaluation examines the correlation between the minimization of CRs and the maximization of aggregation level. The second part presents details about the DRC choices made by PlaceRAN. The third part investigates the relationship between multiple objectives pursued by PlaceRAN and several characteristics of the RAN topology. The fourth part of the evaluation confirms the relevance of the three stages of PlaceRAN. In this part, we also evaluate the impact of the number of paths while employing k-shortest paths. Finally, we summarize our main observations of these four parts of the evaluation at the end of the section. 

\subsubsection*{CRs and Aggregation level}\label{crvsagg}

According to Equation~\ref{eq:of1}, the primary objective of PlaceRAN is to find the best trade-off between the minimum amount of CRs ($\Phi_1$) and the maximum amount of grouped RAN VNFs, i.e., the maximum aggregation level ($\Phi_2$). To make them comparable, in this evaluation, we normalized the amount of CRs and the aggregation level using a percentage scale. The percentage of (employed) CRs corresponds to the ratio of CRs running any positive number of virtualized functions from the set $\mathcal{F}^\prime = \{f_2, f_3, f_4, f_5, f_6, f_7, f_8\}$. To compute the percentage of the aggregation level, we assume that the highest achievable value of $\Phi_2$ is given by $|\mathcal{F}^\prime| \times |\mathcal{B}|$, which represents all virtualized functions running in a single CR.





\begin{figure}[h!]
\subfigure[T1 topology]{ 
\includegraphics[width=0.24\textwidth]{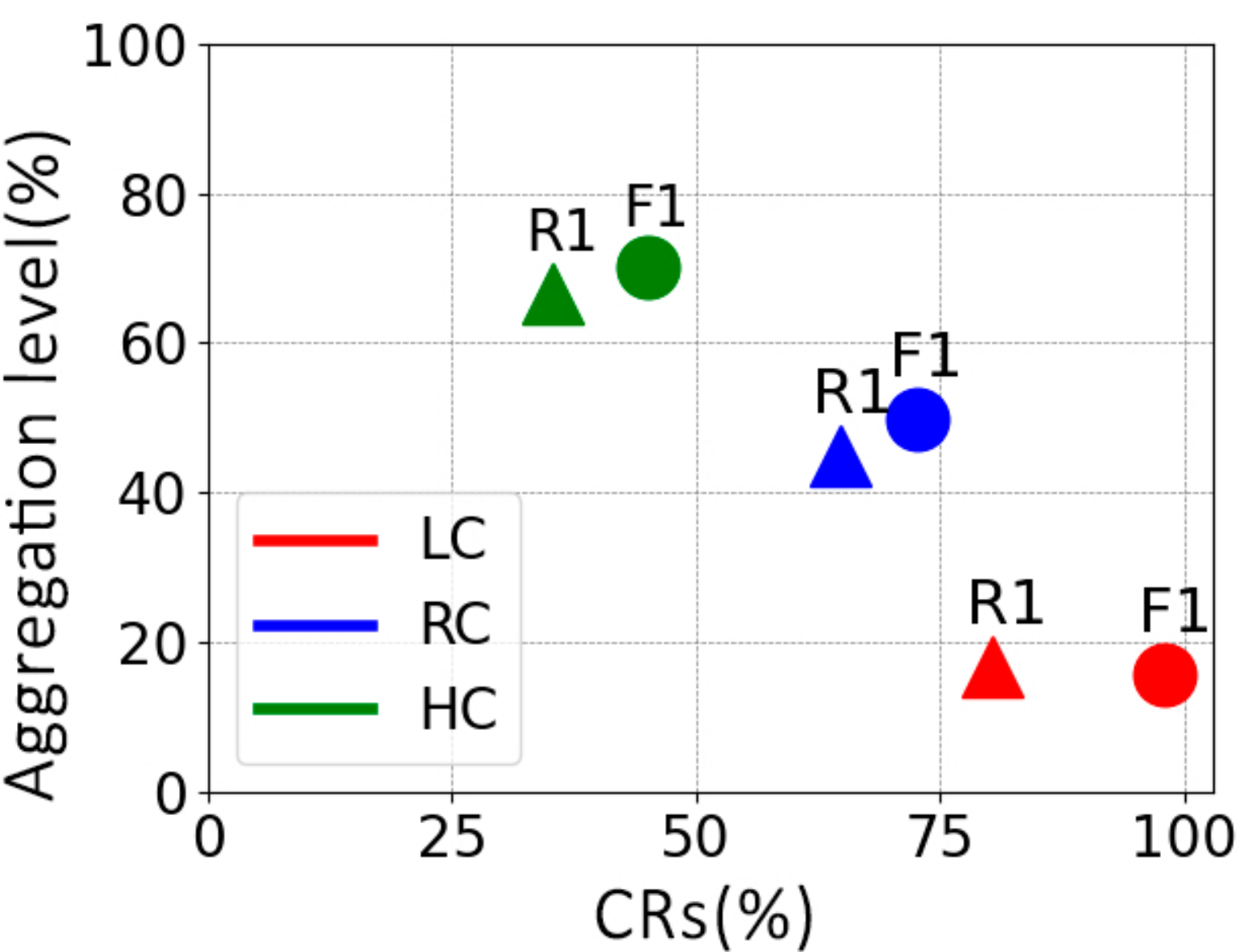}}
\subfigure[T2 topology]{ 
\includegraphics[width=0.24\textwidth]{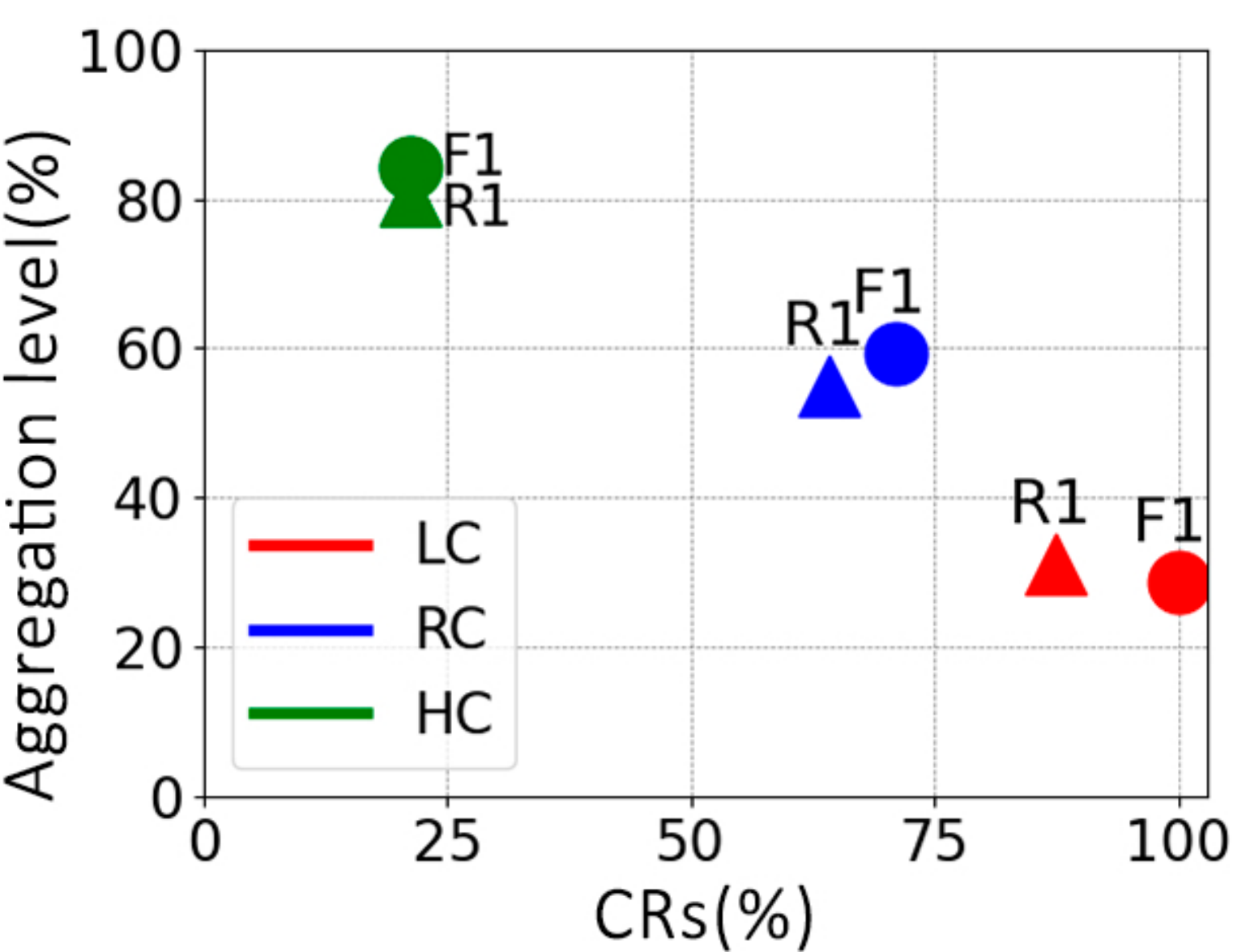}}
\caption{The relation between the percentage of aggregation level and percentage of employed CRs.}
\label{fig:RCvsAGG}
\end{figure}

Fig. \ref{fig:RCvsAGG} shows the percentage of employed CRs (X-axis) and the percentage of aggregation level (Y-axis) of each solution obtained in three scenarios (LC, RC, and HC) and in two configurations of RU nodes (F1 and R1). As expected, as the capacity increases, there is a trend in increasing the aggregation level and decreasing the amount of employed CRs. However, the RAN topology has a significant impact on the solutions. T1 topology, which represents the present RAN topologies, tends to limit the benefits of increasing the resources. For example, when comparing the low amount of resources (LC) with the intermediary amount of resources (RC), the configuration F1 exhibits improvement in aggregation level but not in the number of employed CRs. Additionally, the hierarchical topology adopted by T2 is robust to the configurations of RU nodes, i.e., to the demand, mainly as the amount of resources increases. In general, the modern T2 topology also presents better solutions in both aspects, percentage of aggregation level and percentage of employed CRs.


To evaluate the PlaceRAN behavior in more detail, we present, in Table~\ref{tab:Agg}, the percentage of aggregation level and the number of employed CRs connected to each type of transport node. This table concerns T1 topology and so the types of transport nodes are AG1, AG2, and AC1. The results confirm the need for proper planning of the network resources, more specifically, the link capacity. For example, the aggregation level consistently increases in the CRs connected to AG1 as the network resources increases, i.e., from LC to HC. The results in Table~\ref{tab:Agg} also illustrate that the PlaceRAN model achieves the desired behavior as described in the following. When the network resources are scarce (i.e., LC scenario), PlaceRAN employs more CRs (i.e., a total of 15 for R1 configuration and 22 for F1 configuration) since this approach is the best strategy to increase the aggregation level. As the network resources increase, PlaceRAN employs less CRs, e.g., in the RC scenario, a total of seven CRs is employed in the R1 configuration and 11 CRs in the F1 configuration. In the HC scenario, only three CRs are used in the R1 configuration and only four in the F1 configuration. Additionally, PlaceRAN always seeks to increase the aggregation of the VNFs in the CRs connected to the transport nodes with the best topological conditions, i.e., the AG1 nodes. These nodes are known as ``bridging nodes'' in network science, i.e., nodes connecting densely connected nodes in the network~\cite{ramanathan:06}. In practice, these nodes receive special attention from the infrastructure operators, such as regular maintenance and high security, which confirms the best strategy adopted by PlaceRAN. Finally, Table~\ref{tab:Agg} confirms the relevance of proper dimensioning the network resources to avoid underutilization of the computing resources. For example, the number of employed CRs connected to AG1 does not vary according to the increase in capacity (from LC to HC). This number of employed CRs remains in two, but the aggregation level increases noticeably (e.g., in the F1 configuration, from 18.4\% to 83.7\%) and, as a consequence, also increases the utilization level of these CRs.

\begin{table}[!h]
    \caption{Aggregation level of the T1 topology.}
  \label{tab:Agg}%
    \centering{ {
        \begin{tabular}{|lc|c|c|c|c|c|c|}
        \hline
        \multicolumn{2}{|c|}{} & \multicolumn{2}{c|}{LC} & \multicolumn{2}{|c|}{RC} & \multicolumn{2}{c|}{HC} \\ \cline{3-8}
        
         \multicolumn{2}{|c|}{} & R1 & F1 & R1 & F1 & R1 & F1 \\ \hline
         
         \rowcolor[HTML]{cdd5e4} & * & 26.7 & 18.4 & 58.9 & 53.1 & 92.3 & 83.7 \\ \cline{3-8}
         \rowcolor[HTML]{cdd5e4} \multirow{-2}{*}{AG1} & \# & 2 & 2 & 2 & 2 & 2 & 2 \\ \hline
         
          & * & 10.5 & 0 & 25.7 & 18.4 & 0 & 16.3 \\ \cline{3-8}
          \multirow{-2}{*}{AG2} & \# & 1 & 0 & 1 & 2 & 0 & 2 \\ \hline 
         
         \rowcolor[HTML]{cdd5e4} & * & 62.8 & 81.6 & 15.4 & 28.5 & 7.7 & 0 \\ \cline{3-8}
         \rowcolor[HTML]{cdd5e4} \multirow{-2}{*}{AC1} & \# & 12 & 20 & 4 & 7 & 1 & 0 \\ \hline 
         
       \end{tabular}
        \\ * Aggregation level (\%); \# Number of employed CRs 
    } }
\end{table}

\subsubsection*{DRCs options}
\label{drcdeploy}

As described in Section~\ref{sec:concepts}, each functional split may promote a set of benefits when the VNFs are aggregated. PlaceRAN must also concern about the selection of DRCs since this affects the overall performance of vRAN. Initially, we group the DRCs in four sets according to the fundamental standards to investigate how PlaceRAN selects DRCs. \textbf{NG-RAN~(3)}, formed by DRCs 1, 2, 7, and 8, which are implemented over three independent nodes (CU, DU, and RU); \textbf{NG-RAN~(2)}, formed by DRCs 12 and 13, which are implemented over two independent nodes (CU, DU+RU); \textbf{C-RAN}, formed by DRCs 17 and 18, which are implemented over two other independent nodes (CU+DU, RU); and \textbf{D-RAN}, formed by only DRC19 that is implemented over a single node (CU+DU+RU). Again, we present the results for both topologies (T1 and T2), the capacity scenarios (LC, RC, and HC), and both configurations of RU nodes (F1 and R1).

\begin{figure}[h!] 
\subfigure[T1 topology]{ 
\includegraphics[width=0.238\textwidth]{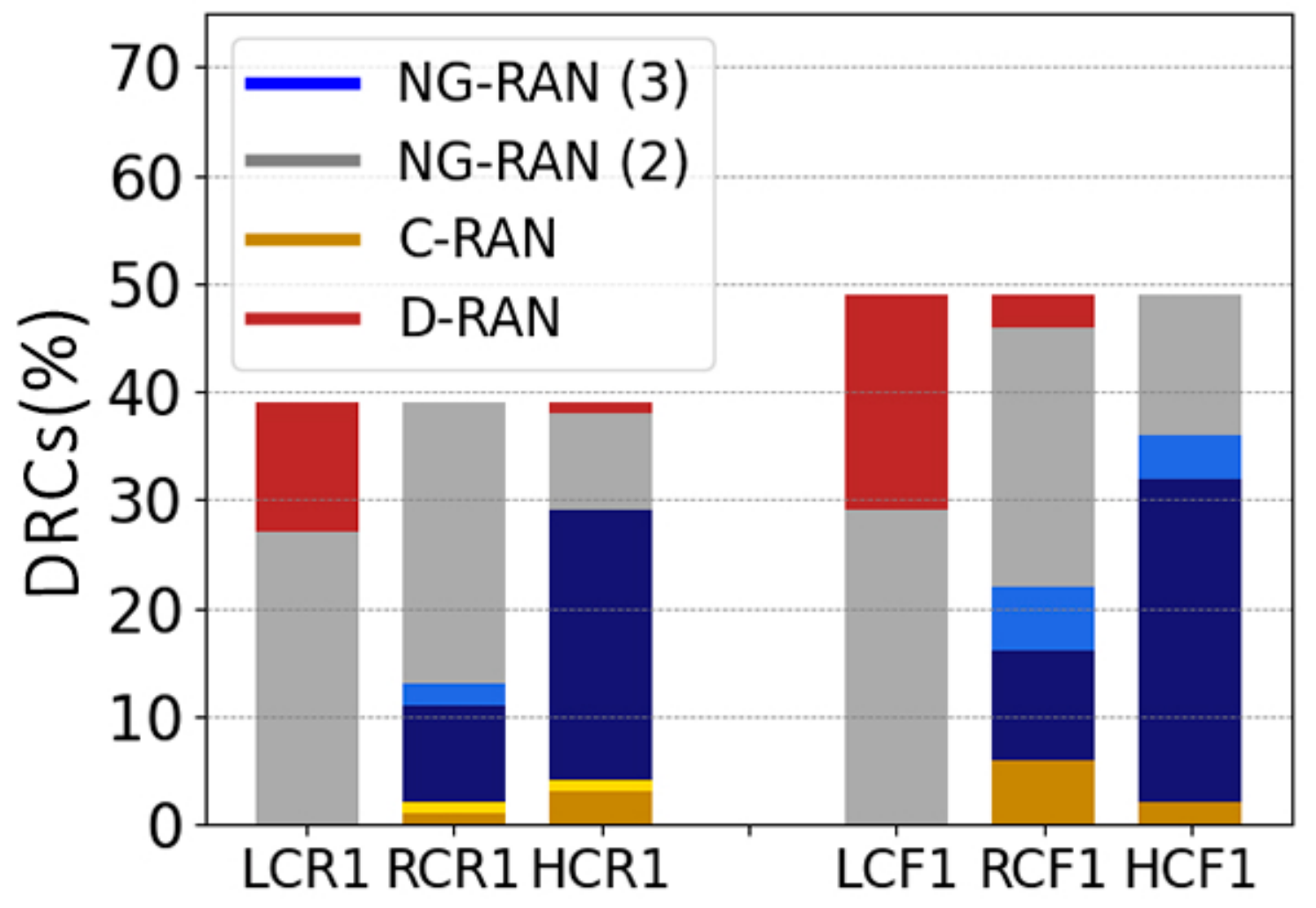}}
\subfigure[T2 topology]{ 
\includegraphics[width=0.246\textwidth]{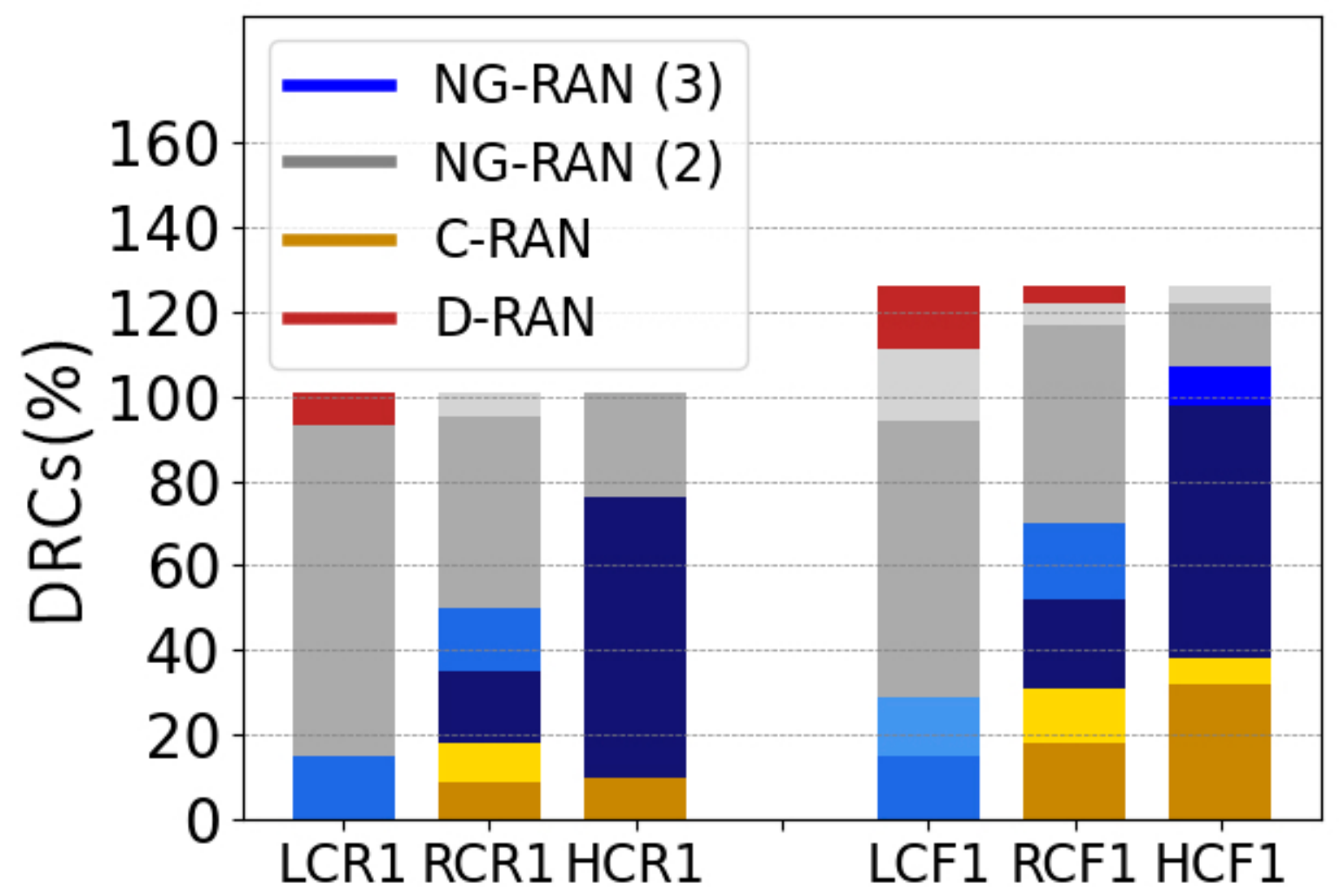}}
\caption{Number of DRCs selected for each scenario and configuration.}
\label{fig:DRC-Usage}
\end{figure}




In Fig.~\ref{fig:DRC-Usage}, each bin represents the number of DRCs selected for each scenario (e.g., LC, RC, or HC) and each configuration (e.g., R1 or F1). We assigned a color to each DRC set (e.g., blue to NG-RAN (3)) and a color shade to each (unique) DRC priority, using Table~\ref{tab:profiles} as reference. The brightness level is assigned in decreasing order from the lowest to the highest DRC priority, e.g., to DRC7 (priority six) is given the brightest blue, while DRC2 (priority one) is assigned the darkest blue. As described in Section~\ref{sec:concepts} and Table~\ref{tab:profiles}, the DRC sets can be ranked as follows. C-RAN and NG-RAN (3) are preferable, but C-RAN presents high demand in the fronthaul and, so its deployment tends to be limited. NG-RAN (2) is the next preferable choice, offering several advantages against D-RAN, which brings few benefits in a vRAN context. As a general observation, PlaceRAN follows a strategy aligned with this ranking. In both topologies (T1 and T2) and configurations of RU nodes (F1 and R1), as the amount of resources increases (from LC to HC), PlaceRAN tends to decrease the amount of D-RAN and increases the amount of C-RAN and NG-RAN (3). Some additional observations are presented next.

While D-RAN is the least preferable DRC for vRAN, it may appear as part of an optimal solution to achieve the best global trade-off between aggregation level and the number of CRs. This situation explains why HCR1 has D-RAN while RCR1 does not have, as shown in Fig.~\ref{fig:DRC-Usage}. On the other hand, HCR1 exhibits a noticeably increase in NG-RAN (3) and C-RAN. Again, T2 topology confirms the benefits of the hierarchical approach since the improvement in selecting the DRCs sets is very consistent with the increase of the network resources. In addition to the NG-RAN (3) set, T2 topology makes feasible and beneficial the demanding C-RAN set for several RU nodes, as illustrated in RCF1 and HCF1.

\subsubsection*{Model objectives and RAN topology characteristics}
\label{comparison}

In this part of the evaluation, we analyze some outputs from the PlaceRAN model: aggregation level from Stage 1, number of (unique) DRCs from Stage 2, and the sum of DRC priorities from Stage 3. Additionally, we analyze several RAN characteristics: average network latency, average occupation of CRs connecting to each type of node, and average link occupation of each type of node. Similar to our previous approach, we normalize all the values using a percentage scale, as described in the following. 

\begin{figure}[h!]
\centering
\subfigure[T1 topology]{ 
\includegraphics[width=0.235\textwidth]{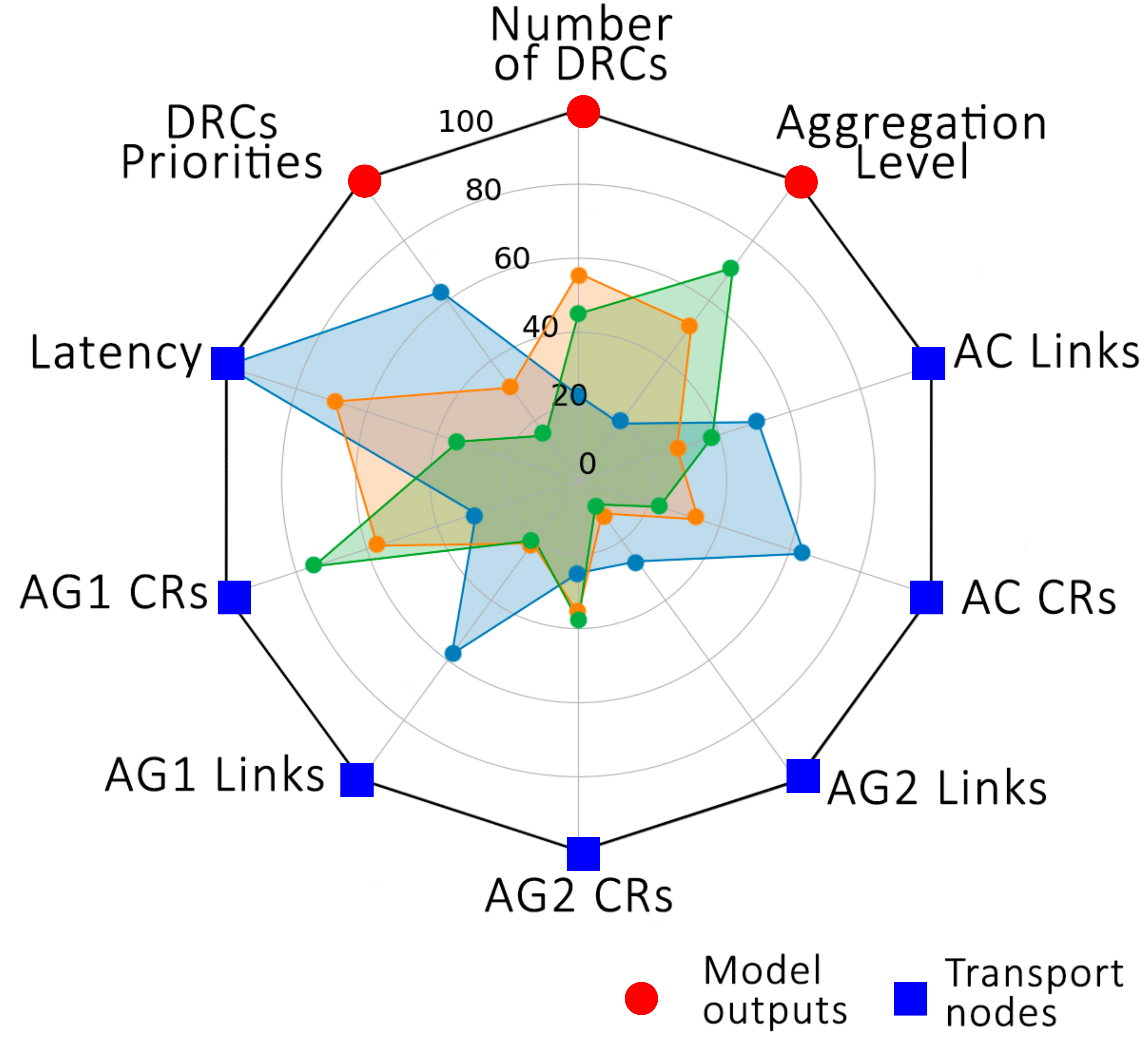}}
\label{fig:graph_radar_tim_R1}
\subfigure[T2 topology]{ 
\includegraphics[width=0.235\textwidth]{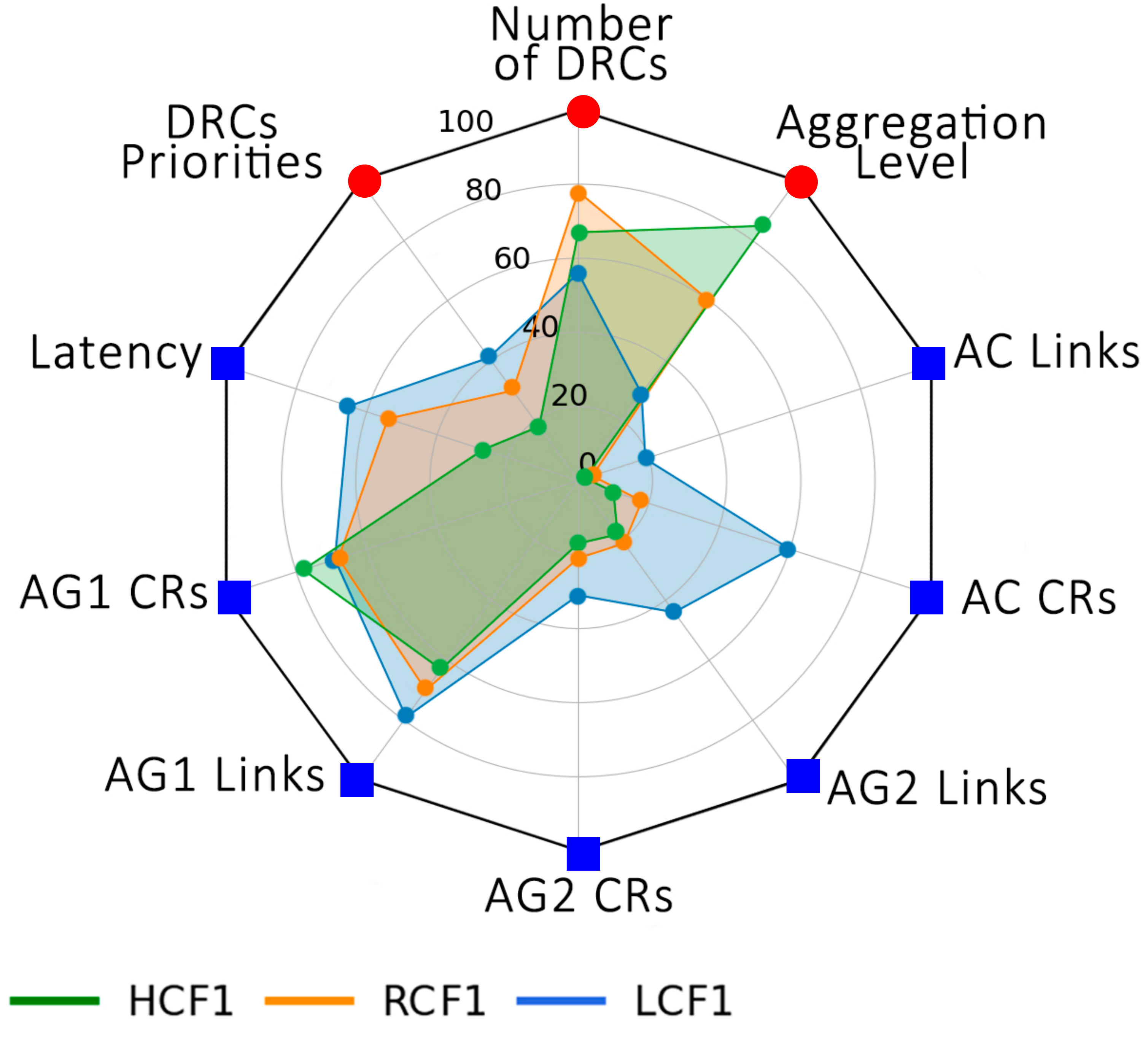}}
\label{fig:graph_radar_tim_F1}
\caption{Relationship between some PlaceRAN outputs and several RAN characteristics. In T1 topology, AC corresponds only to CR1, while in T2 topology, AC corresponds to AC1 and AC2, as shown in Fig.~\ref{fig:topologies}.}
\label{fig:Radar}
\end{figure}

The percentage of aggregation level is computed as previously, i.e., assuming that the highest achievable value of $\Phi_2$ is given by $|\mathcal{F}^\prime| \times |\mathcal{B}|$. The percentage of the number of DRCs is obtained by taking nine as the largest possible number of DRCs (as described in Table~\ref{tab:splits}). To compute the percentage of the sum of DRC priorities, the highest possible value is obtained by taking the product of the number of RUs ($|\mathcal{B}|$) and the priority of D-RAN, which has the highest value. Therefore, the worst case, in which all RUs would use D-RAN, corresponds to 100\% in the sum of DRC priorities. The average network latency, computed over all links, of the low capacity scenario from T1 topology is taken as the maximum value, and from this, the percentage of average network latency is computed accordingly for all scenarios in both topologies. The percentage of average occupation of CRs connecting to each type of node (AG1, AG2, or AC) is effectively the measurement of the percentage of the average amount of utilized computing resources. As previously described, each CR is connected to a type of node, i.e., AG1, AG2, or AC. Therefore, the average is computed over all CRs connecting to a certain type. Similarly, the percentage of average link occupation of each type of node is measured over all links connected to all nodes of a certain type. These previous two measurements (CR and link occupation) are obtained directly from the solution provided by PlaceRAN, i.e., after the routing decisions and placement of VNFs. Fig.~\ref{fig:Radar} present these results, which are discussed in the following.

The aggregation level and sum of DRC priorities are strongly correlated with the average latency, as expected. The average latency decreases as the network resources (i.e., links) increase and both improvements (in latency and link capacity) allow the increase in the aggregation level and the sum of DRC priorities in both topologies. In general, T2 topology has lower average latency than T1 topology and a larger average link capacity (as shown in Table~\ref{tab:evalparam2}), which contributes to clarifying the better performance of T2 topology in the selection of DRCs sets (shown in Fig.~\ref{fig:DRC-Usage}). On the other hand, the number of (unique) DRCs does not exhibit such a clear relationship with any RAN characteristic. While the benefit of decreasing the number of (unique) DRCs is quite clear, its relationship with the other model outputs is not trivial and is affected by RAN characteristics.

Fig.~\ref{fig:Radar} also illustrates the relevance of planning joint changes in the network resources and deployment of computing resources. In general, increase the network resources promotes an increase in the utilization of CRs connecting to AG1 nodes and a decrease in the utilization of CRs connecting to AC nodes since this tends to be the best strategy to improve the aggregation level. However, the link occupation of these nodes exhibit a more elaborated behavior, which is clearly influenced by the topology. For example, in the T2 topology, increasing the network resources implies decreasing the utilization of AG1 and AC links, while in the T1 topology, there is not this consistent behavior. This behavior happens because, in the T1 topology, PlaceRAN finds a better trade-off solution by employing CRs connecting to AG2 nodes. In addition to vRAN, this knowledge can also contribute to the planning for edge computing adoption.

\subsubsection*{Stages of PlaceRAN model and number of paths}
\label{modelassess}

Previously, in Section~\ref{sec:model}, we provided simple examples that illustrate the need for each of the three stages of PlaceRAN. While those examples can justify the model design, they do not show the effective impact of this approach in real-world instances. To show this impact, in Table~\ref{tab:Stages}, we present the selection of DRC sets in each PlaceRAN stage. These results refer to T1  topology, R1 configuration of RU nodes, and RC scenario, i.e., an intermediate amount of resources. Each stage exhibits a clear difference in the selected DRC sets, confirming, in the last stage, the preference for NG-RAN (3) or C-RAN and the avoidance of D-RAN, if possible. Moreover, as previously discussed, eventually, all stages may present the same selection of DRC sets. However, we observed changes in the selection of DRC sets along the stages in all scenarios, configurations, and both topologies.



\begin{table}[!h]
    \caption{Analyzing the three stages of the PlaceRAN model.}
  \label{tab:Stages}%
    \centering{ {
        \begin{tabular}{|l|c|c|c|}
        \hline
        DRCs & Stage 1 & Stage 2 & Stage 3 \\ \hline
        
        \rowcolor[HTML]{cdd5e4} 1 - NG-RAN(3) & 0\% & 0\% & 0\% \\ \hline

         2 - NG-RAN(3) & 17.9\% & 20.5\% & 23.1\% \\ \hline 
        
        \rowcolor[HTML]{cdd5e4}  7 - NG-RAN(3) & 0\% & 0\% &  0\% \\ \hline
        
          8 - NG-RAN(3) & 5.1\% & 10.2\% & 5.1\% \\ \hline

        \rowcolor[HTML]{cdd5e4}   12 - NG-RAN(2) & 0\% & 0\% & 0\% \\ \hline
        
          13 - NG-RAN(2) & 61.4\% & 64.1\% & 66.6\% \\ \hline
         
         \rowcolor[HTML]{cdd5e4}  17 - C-RAN & 5.2\% & 0\% & 2.6\% \\ \hline
         
          18 - C-RAN & 5.2\% & 2.6\% & 2.6\% \\ \hline
         
         \rowcolor[HTML]{cdd5e4}  19 - D-RAN & 5.2\% &  2.6\% &  0\%  \\ \hline

        \rowcolor[HTML]{3c4563} \color{white} Total & \color{white} 100\% & \color{white} 100\% & \color{white} 100\%  \\ \hline
       \end{tabular}
    } }
\end{table}

PlaceRAN, similarly to other optimization problems that need to compute optimal routes, is affected by the number of paths available as input. Theoretically, all paths available in a graph should be provided as input to the model. However, in real-world problems, this approach is rarely adopted due to some reasons. First, even a small graph (or topology), e.g., with 51 nodes, may have a huge number of paths, making the computation time unacceptable. Second, even removing the loops, many paths are commonly useless because they do not satisfy constraints such as latency or capacity. Therefore, in general, a subset of the available paths is provided as input and a common choice is the k shortest paths. Since this approach is employed to provide input to PlaceRAN, we were interested in understanding the impact of the k value in the solutions provided by our model.



\begin{figure}[h!]
\centering
\includegraphics[width=0.48\textwidth]{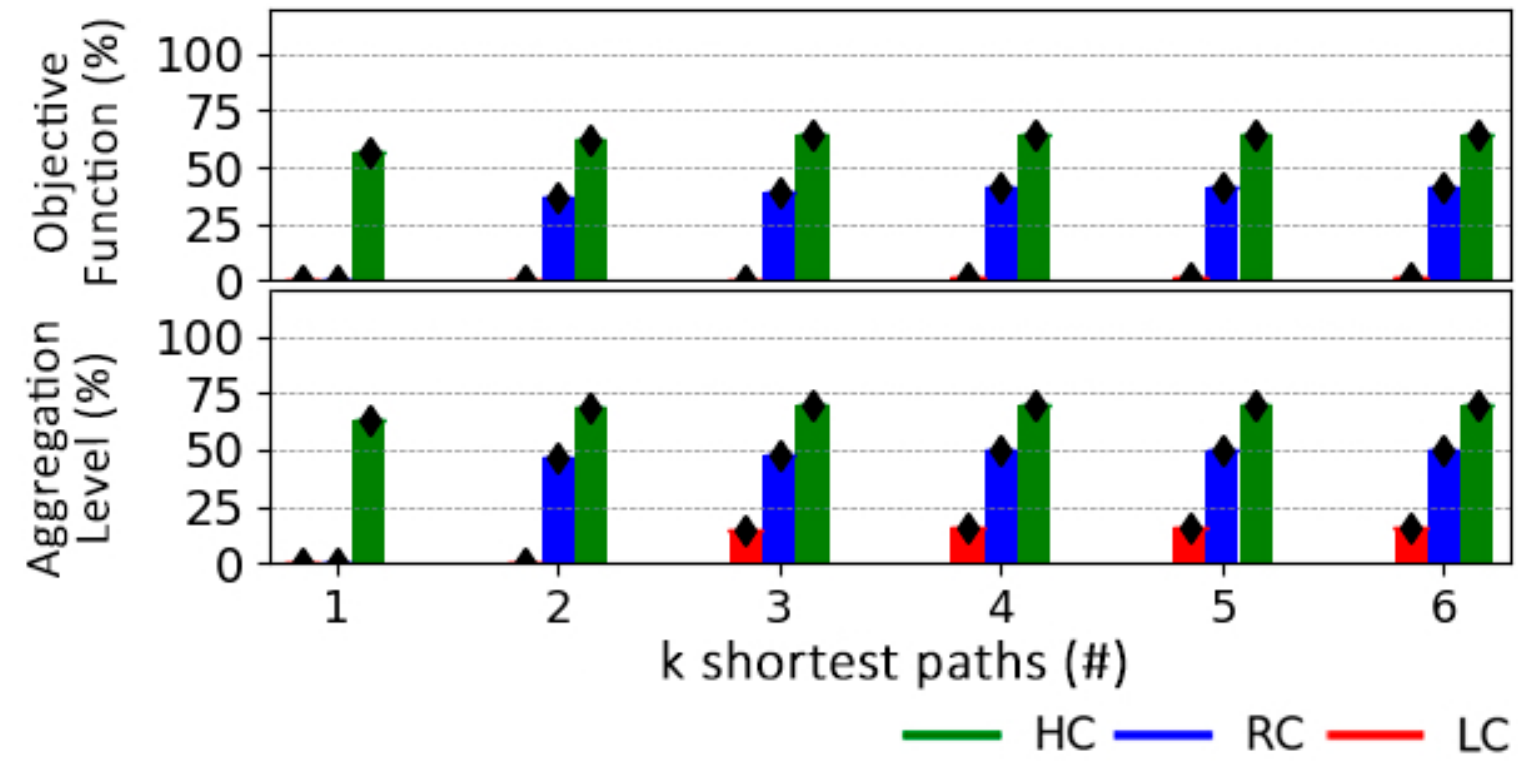}
\label{fig:k_paths_TIM}
\caption{K shortest paths for the T1 topology and F1 configuration of RU nodes.}
\label{fig:k_paths}
\end{figure}

In PlaceRAN, each set of k shortest paths is generated having as endpoints an RU and the core. Again, we normalize all the values using a percentage scale as  follows. The value of the objective function (of the first stage) with all RUs using C-RAN from a single CR is considered as 100\%. This behavior is rarely a feasible solution, but it works well as a reference. Similarly, the aggregation level percentage is obtained assuming that all RUs using C-RAN from a single CR corresponds to 100\%. Fig.~\ref{fig:k_paths} presents the results for the T1 topology and F1 configuration of RU nodes. In general, there is no noticeable difference from $k \geq 3$ in both outputs, objective function, and aggregation level. We adopted $k=4$ to generate the paths because this is the largest value possible for the T2 topology since it is not allowed to repeat a hierarchical level in a path (in this topology). With $k \leq 2$, objective function and aggregation level become zero in some scenarios (e.g., LC and RC for $k=1$) because PlaceRAN does not find any solution. From $k \geq 3$, objective function and aggregation level are always higher than zero. However, objective function values for the LC scenario when $k \geq 3$ are very low (i.e., close to zero) because many RUs use D-RAN (as illustrated in Fig.~\ref{fig:DRC-Usage}).

\subsubsection*{Discussion}


Initially, we highlight the efficiency of PlaceRAN, which optimally solved the problem for real-world RAN instances up to 128 nodes. Through the thorough evaluation, we also confirmed the suitability of PlaceRAN to represent the problem of finding the best trade-off between the aggregation level and the number of computing resources, placing optimally the VNFs of a vRAN under network and computing constraints. Given the complexity of this trade-off, in addition to the overall analyses of results, we performed a detailed investigation by selecting some solutions, mainly the nontrivial ones. For example, in Fig.~\ref{fig:DRC-Usage}, there is D-RAN in an HC scenario while the RC scenario does not have in the RC scenario (in T1 topology, R1 configuration of RU nodes). After investigating this example, we observed the solution provided by PlaceRAN was effectively the best trade-off, and we were not able to improve it by any modification, as expected. All the investigated solutions were shown to be aligned with the desired behavior.

In the evaluation, we also identified some relevant insights related to vRAN in a broad sense. For example, to support the demand of 5G and beyond networks, in which RUs must provide several Gbps and assure very low latency, the traditional RAN topologies (represented by T1 in our evaluation) are not appropriate. The hierarchical approach proposed by the PASSION project (represented by T2 in our evaluation) seems in the right direction since it exhibited superior performance in comparison with the traditional approach. Even with a larger demand from the RUs, in T2, PlaceRAN is able to select a higher number of NG-RAN (3) and C-RAN DRCs.

Finally, although the work does not explore the cost issue, the three capacity scenarios chosen are directly related to cost. For example, in the HC scenario, which exhibits the best solution performance, there is a large quantity of expensive resources in the network and consequently higher investment. In both topologies, T1 based on the 5G-crosshaul project and T2 based on the PASSION project, we observed an imbalance in the occupation of the resources. Using again the HC scenario as an example, AC nodes presented low occupation in links and in the connected CRs, which can be optimized, consequently reducing investment.

\section{Related Work}
\label{rw}

\begin{table*}[!h]
    \caption{Related work.}
    \label{tab:related}
    \centering{ {
        \begin{tabular}{|l|l|c|c|}
        \hline
        \multicolumn{1}{|c|}{} & \multicolumn{1}{c}{Optimization} & \multicolumn{2}{|c|}{Disaggregated RAN}  \\ \cline{3-4} 
        \multirow{-2}{*}{Works}   & \multicolumn{1}{c}{Goals} & \multicolumn{1}{|c}{Nodes}  & \multicolumn{1}{|c|}{DRCs} \\ \hline \hline

        \rowcolor[HTML]{cdd5e4} \multicolumn{1}{|c|}{\cite{garcia2018fluidran}} & Maximization of CU centralization and evaluation of Latency Edge & CU-RU & 3 \\
    
        \multicolumn{1}{|c|}{\cite{fonseca:19}} & Maximization of CU centralization and flexible CU positioning & CU-RU & 5 \\
        
        \rowcolor[HTML]{cdd5e4}
        \multicolumn{1}{|c|}{\cite{bhamare2018efficient}} & Maximization of BBU centralization and minimize the overall latency &
        BBU-RRH & 1  \\

        \multicolumn{1}{|c|}{\cite{harutyunyan2020cu}} & Dynamic CU and wireless fronthaul to minimize the energy & CU-DU & 4 \\
        \rowcolor[HTML]{cdd5e4}
        \multicolumn{1}{|c|}{\cite{arouk2017multi}} & Minimize the cost of VNF chain on the substrate network & BBU-RRH & 1 \\
        
        \multicolumn{1}{|c|}{\cite{mahapatra2017optimal}} & Identify the ideal position of the centroid for the BBU & BBU-RRH & 1   \\
        
        \rowcolor[HTML]{cdd5e4}
        \multicolumn{1}{|c|}{\cite{molner2019optimization}} & Maximize DU distribution under a non-dedicated optical network & CU-DU & 2  \\
        
        \multicolumn{1}{|c|}{\cite{matoussi20205g}} & Minimize the CRs and fronthaul bandwidth & BBU-RRH & 3 \\
        
        \rowcolor[HTML]{cdd5e4}
        \multicolumn{1}{|c|}{\cite{song2019clustered}} & Minimize the computational costs & Monolithic & 1  \\
        
        \multicolumn{1}{|c|}{\cite{arouk2018cost}} & Minimize the CRs and overall routing costs &     CU-DU-RU & 1    \\
        
        \rowcolor[HTML]{cdd5e4}
        \multicolumn{1}{|c|}{\cite{masoudi2020cost}} & Minimize the total cost of ownership by DU pool & DU-RU & 4     \\
        
         \multicolumn{1}{|c|}{\cite{yusupov2018multi}}  & Minimize the nodes and latency; maximize the data rate & CU-DU-RU & 2      \\
        
        \rowcolor[HTML]{cdd5e4} 
        \multicolumn{1}{|c|}{\cite{harutyunyan2018flex5g}} & Minimize interference and fronthaul links to optimize the network & CU-DU & 4   \\
        
        \multicolumn{1}{|c|}{\cite{murti2020optimization}, \cite{murti:20}} & Minimize the vRAN cost and overall routing, based on CU's positioning  & CU-DU-(RU) & 3  \\
        \rowcolor[HTML]{3c4563} \multicolumn{1}{|c|}{\color{white} PlaceRAN}  &  \color{white} Minimize CRs and maximize vNG-RAN's radio functions aggregation & \color{white} CU-DU-RU & \color{white} 9 \\  \hline

       \end{tabular}
    } }
\end{table*}

Recently, RAN has faced an intense process of \textit{softwarization} and virtualization, which has driven into a fast evolution. However, this scenario also created a misalignment of several works related to the (virtual) network function placement due to the raise of multiple initiatives in a short period of time. These initiatives are usually led by the standard development organizations, e.g., 3GPP. Moreover, the O-RAN alliance drives some directions, such as vNG-RAN focusing on the disaggregation of radio functions for network efficiency and performance. In this context, Table \ref{tab:related} shows several relevant works regarding Optimization Goals and Disaggregated RAN, considering the type of the nodes modeled and the number of DRCs.



\textbf{Optimization Goals} - this characteristic is the most relevant when comparing the related work since investigations with different optimization goals have distinct problem formulations, achieve different results, and, commonly, drive to diverse insights. In some works, the focus is to maximize the number of VNFs running in a single CU. For example, for Garcia-Saavedra et al. \cite{garcia2018fluidran}, the CU is co-located with the core, and D-RAN is not an option. However, for Fonseca, Correa, and Cardoso \cite{fonseca:19}, D-RAN is considered, and different positions are evaluated for CU. The maximization of CU and BBU nodes are aimed at several investigations \cite{garcia2018fluidran,bhamare2018efficient,mahapatra2017optimal,fonseca:19,harutyunyan2018flex5g}. In C-RAN architecture, the fronthaul latency and the capacity of data rate links are widely analyzed \cite{harutyunyan2020cu,arouk2017multi,matoussi20205g,yusupov2018multi,harutyunyan2018flex5g,murti2020optimization}. Based on RAN centralization, Song et al. \cite{song2019clustered} and Matoussi et al. \cite{matoussi20205g} focus on the CRs efficiency. Furthermore, the RAN transformation is targeted into the maximization of centralization, the CRs efficiency, and crosshaul link cost assessment by Arouk et al. \cite{arouk2018cost} and Masoudi, Lisi, and Cavdar \cite{masoudi2020cost}. 

Murti et al., in their initial work \cite{murti2020optimization} and in its extended version \cite{murti:20}, have the closest investigation to ours. These authors are also approaching the problem of finding the best trade-off between the minimum number of CRs and maximum aggregation level. However, they consider only vCUs, while DUs are fixed and close to (RU)s, i.e., provided as input. In our work, we consider not only vCUs, but also vDUs, making our problem more general. We formulated the problem as a Nonlinear Programming model, with binary variables, linear constraints, and a nonlinear objective function (from the first stage). We were able to create an equivalent representation of the nonlinear objective function using Min and Max functions. Therefore, our model can be solved exactly by a \textit{conventional} solver, e.g., the IBM CPLEX.

\textbf{Disaggregated RAN} - the number of types of the RAN nodes and the number of DRCs are related to the flexibility and complexity of the problem under investigation. The types of the RAN nodes available and the number of DRCs represented are related to the accuracy of describing the real-world disaggregated RANs. In this context, our model represents all types of the RAN nodes and all industry DRCs, i.e., our model describes precisely the present disaggregated RANs. Additionally, our model is ready for supporting other DRCs by just changing the set $\mathcal{D}$. 

We adopted a conservative approach during the classification of the related work in terms of RAN nodes and the number of DRCs.
Therefore, most investigations had their benefits amplified for the disaggregated RAN, e.g., Murti et al. \cite{murti2020optimization, murti:20} describe CU, DU, and RU in the system model. Furthermore, the paths between RU are not taken into account.
This situation happens because each RU has only a single link connecting to a single DU. Therefore, our model is again more general than the state-of-the-art.



Given the complexity of the VNF placement problem, several works in the literature adopt an approximate and heuristic approach \cite{bhamare2018efficient,harutyunyan2020cu,arouk2017multi,molner2019optimization,matoussi20205g,harutyunyan2018flex5g}. As expected, in general, the main advantage of this type of approach is the reduced computing cost, and the main drawback is the suboptimal solutions. As discussed in~\cite{murti:20}, suboptimal vRAN solutions have a large cost impact in the long-term. Additionally, our strategy has advantages in comparison with approximate and heuristic approaches. First, PlaceRAN is able to obtain the optimal solution in satisfactory time for several real-world networks, mainly for the most modern topologies. Second, when the computation is interrupted before achieving the optimum, we know the gap of the suboptimal solution obtained by PlaceRAN.

\section{FINAL REMARKS}
\label{conc}

In this work, we have formulated the vNG-RAN placement problem and introduced the Disaggregated RAN Combination (DRC) concept. This problem is still a challenge for academia and industry for deploying the fifth-generation RAN. We designed an optimal solution to deliver the best possibility of allocating RAN's VNFs to reduce computational resources and reach maximum aggregation and consequent centralization of RAN's protocols and units. 
Additionally, we choose the best path to overcome the limitations of crosshaul networks. We reduced the number of DRCs in the network and inserted the strategy concept for choosing DRCs. The scenarios analyzed were based on real networks with the possibility of implementation in any vNG-RAN.


For future work, we envision advances in the solution concerning the time factor. We will evolve the solution to deal with traffic in the flow format, avoiding bandwidth waste in the scope of choosing the best path for crosshaul networks. Moreover, our strategy can be more aligned with O-RAN initiatives considering computing resources. Therefore, we will be introducing the type of processing to be defined for each virtualized radio function, e.g., general-purpose processor (GPP) or specific purpose processor (SPP).


\section*{Acknowledgment}

This work was conducted with partial financial support from the National Council for Scientific and Technological Development (CNPq) under grant number 130555/2019-3 and from the Coordination for the Improvement of Higher Education Personnel (CAPES) - Finance Code 001, Brazil.

\bibliographystyle{IEEEtran}
\bibliography{biblio}

\begin{IEEEbiography}[{\includegraphics[width=1in,height=1.1in,clip]{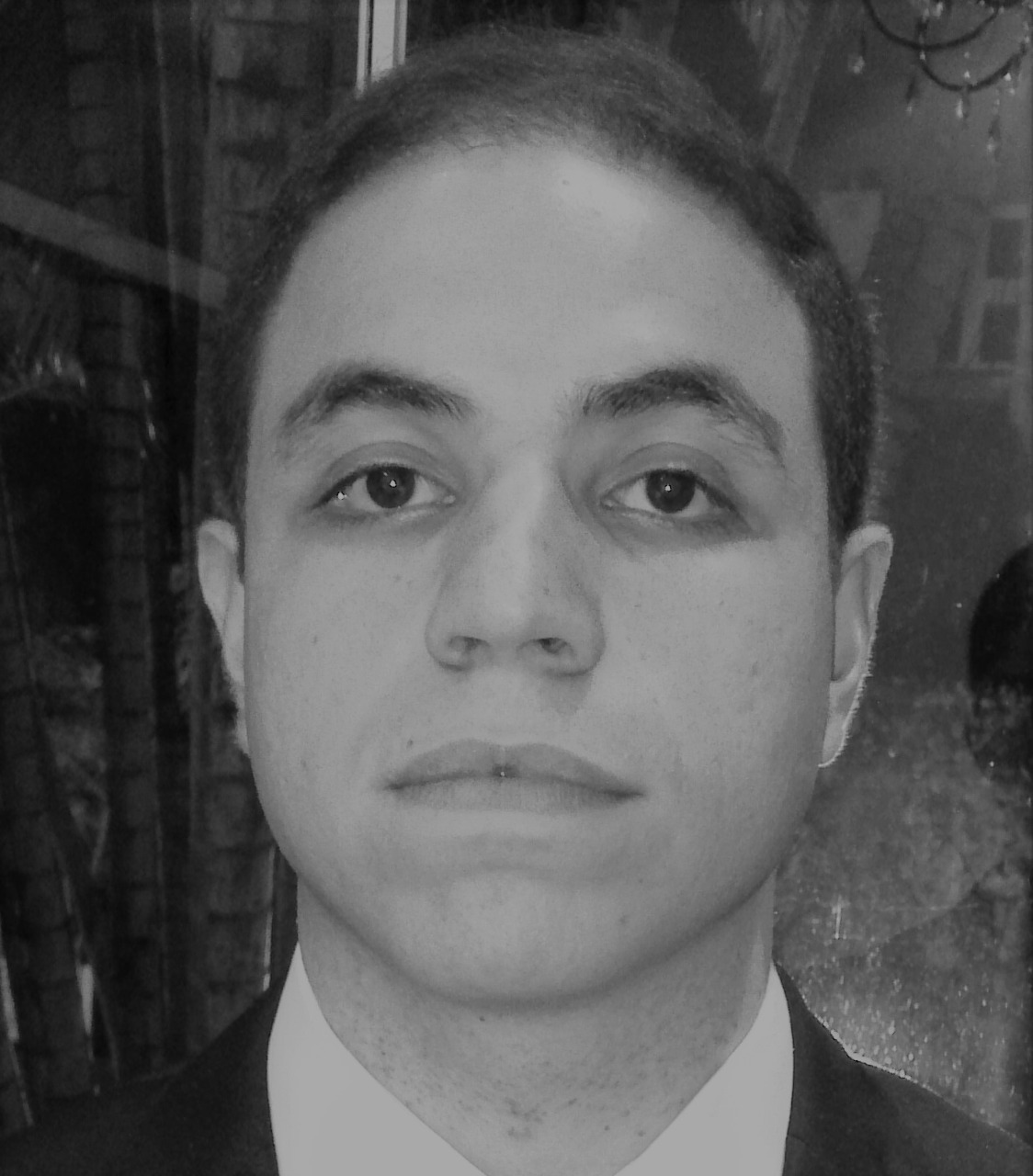}}]{Fernando Zanferrari Morais} is MSc degree at University of Vale do Rio dos Sinos (UNISINOS), Brazil. Fernando received the bachelor's degree from Centro Universitário Lasalle in 2009. He served as a Telecommunication Engineer at Vivo Operator, a Telefonica group, Brazil. His research interests include telecommunication mobile networks, software-defined networking, virtualization and cloud computing. \end{IEEEbiography}

\vspace{-1.5cm}

\begin{IEEEbiography}[{\includegraphics[width=1in,height=1.1in,clip]{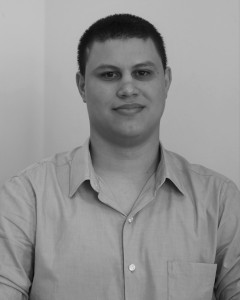}}]{Gabriel Matheus F. de Almeida} has been a Computer Science Researcher and a member with the Laboratory Computer Networks and Distributed Systems, Federal University of Goiás (UFG), Brazil, since 2019. His research interests include wireless networks, software-defined networks, virtualization, resource allocation and performance evaluation. \end{IEEEbiography}
 
 \vspace{-1.5cm}

\begin{IEEEbiography}[{\includegraphics[width=1in,height=1.1in,clip]{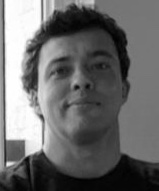}}]{Leizer Pinto} is associate professor at Federal University of Goiás, Brazil. B.Sc., Computer Science, Pontifical Catholic University of Goiás, Brazil, 2004. M.Sc., 2007, and D.Sc., 2009, Systems Engineering and Computer Science, COPPE/UFRJ, Brazil. Visiting PhD Student, 2009, HEC-Montral, Canada. Research interests include Multi-objective Combinatorial Optimization and Network Optimization. \end{IEEEbiography}

\vspace{-1.5cm}

\begin{IEEEbiography}[{\includegraphics[width=1in,height=1.1in,clip]{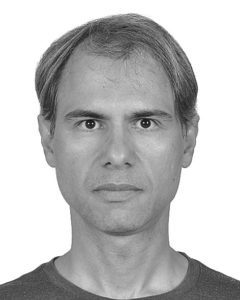}}]{Kleber Vieira Cardoso} is an associate professor at the Institute of Informatics – Universidade Federal de Goiás (UFG), where he has been a professor and researcher since 2009. He holds a degree in Computer Science from UFG (1997), has MSc (2002) and PhD (2009) in Electrical Engineering from COPPE – Universidade Federal do Rio de Janeiro. In 2015, he spent his sabbatical at Virginia Tech (in the USA) and, in 2020, at Inria Saclay Research Centre (in France). \end{IEEEbiography}

\vspace{-1.5cm}

\begin{IEEEbiography}[{\includegraphics[width=1in,height=1.1in,clip]{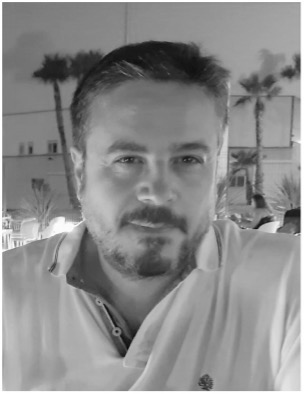}}]{Luis M. Contreras} earned a Telecom Engineer (M.Sc.) degree at the Universidad Politécnica of Madrid (1997), and holds an M.Sc. on Telematics from the Universidad Carlos III of Madrid (2010). Since August 2011 he is part of Telefónica I+D / Telefónica CTIO unit, working on 5G, SDN, virtualization, and transport networks. He is part-time lecturer at the Universidad Carlos III of Madrid, where is also pursuing a Ph.D. (expected for 2021).  \end{IEEEbiography}

\vspace{-1.5cm}

\begin{IEEEbiography}[{\includegraphics[width=1in,height=1.1in,clip]{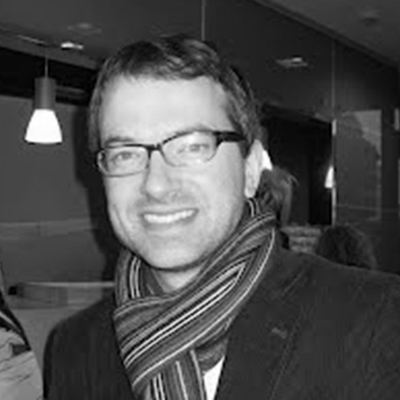}}]{Rodrigo da Rosa Righi} is professor and researcher at University of Vale do Rio dos Sinos (UNISINOS), Brazil. Rodrigo concluded his post-doctoral studies at KAIST - Korean Advanced Institute of Science and Technology, South Korea. He obtained his MS and PhD degrees in Computer Science from the Federal University of Rio Grande do Sul, Brazil, in 2005 and 2009, respectively. He is a member of both IEEE Computer Society (senior member) and ACM. \end{IEEEbiography}

\vspace{-1.5cm}

\begin{IEEEbiography}[{\includegraphics[width=1in,height=1.1in,clip]{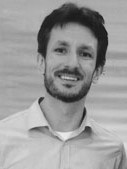}}]{Cristiano Bonato Both} is a professor of the Applied Computing Graduate Program at the University of Vale do Rio dos Sinos (UNISINOS), Brazil. He coordinators research projects funded by H2020 EU-Brazil, CNPq, FAPERGS, and RNP. His research focuses on wireless networks, next-generation networks, softwarization and virtualization technologies for telecommunication networks. \end{IEEEbiography}

\end{document}